\documentclass[superscriptaddress, twocolumn, prl, aps, longbibliography, nofootinbib]{revtex4-1}

\usepackage{amsmath, amssymb, color, wasysym, esint}
\makeatletter
\newsavebox{\@brx}
\newcommand{\llangle}[1][]{\savebox{\@brx}{$\m@th{#1\langle}$}%
  \mathopen{\copy\@brx\kern-0.5\wd\@brx\usebox{\@brx}}}
\newcommand{\rrangle}[1][]{\savebox{\@brx}{$\m@th{#1\rangle}$}%
  \mathclose{\copy\@brx\kern-0.5\wd\@brx\usebox{\@brx}}}
\makeatother

\makeatletter
\newsavebox{\@brxx}
\newcommand{\lllangle}[1][]{\savebox{\@brxx}{$\m@th{#1\langle}$}%
  \mathopen{\copy\@brxx\kern-0.5\wd\@brxx\usebox{\@brxx}\kern-0.5\wd\@brxx\usebox{\@brxx}}}
\newcommand{\rrrangle}[1][]{\savebox{\@brxx}{$\m@th{#1\rangle}$}%
  \mathclose{\copy\@brxx\kern-0.5\wd\@brxx\usebox{\@brxx}\kern-0.5\wd\@brxx\usebox{\@brxx}}}
\makeatother

\usepackage{graphicx}
\usepackage{bm}
\usepackage{xcolor}
\usepackage{xspace}
\usepackage{float}
\usepackage{lineno}

\definecolor{linkcolor}{rgb}{0,0,0.6} 
\usepackage[pdftex,colorlinks=true,
	pdfstartview = FitV,
	linkcolor    = linkcolor,
	citecolor    = linkcolor,
	urlcolor     = linkcolor,
	hyperindex   = true,
	hyperfigures = false]{hyperref}

\begin{document}

\title{Defect statistics in pulsating active liquids: From contraction waves to aster dynamics}

\author{Tirthankar Banerjee}
\email{tirthankar@niser.ac.in}
\affiliation{Department of Physics and Materials Science, University of Luxembourg, L-1511 Luxembourg City, Luxembourg}
\affiliation{School of Physical Sciences, National Institute of Science Education and Research, Jatani, Odisha, 752050 India}
\affiliation{Homi Bhaba National Institute, Training School Complex, Anushakti Nagar, Mumbai, 400094 India}

\author{Luca Cocconi}
\affiliation{Cavendish Laboratory, University of Cambridge, Cambridge CB3 0US, United Kingdom}

\author{Thibault Desaleux}
\affiliation{Department of Physics and Materials Science, University of Luxembourg, L-1511 Luxembourg City, Luxembourg}

\author{Jonas Ranft}
\email{jonas.ranft@ens.psl.eu}
\affiliation{Institut de Biologie de l’ENS, \'Ecole Normale Sup\'erieure, CNRS, Inserm, Universit\'e PSL, 46 rue d’Ulm, 75005 Paris, France}

\author{\'Etienne Fodor}
\email{etienne.fodor@uni.lu}
\affiliation{Department of Physics and Materials Science, University of Luxembourg, L-1511 Luxembourg City, Luxembourg}

\begin{abstract}
We propose and study a hydrodynamic theory that captures the emergence of contraction waves in dense active liquids composed of pulsating particles subject to isotropic deformation. The coupling between displacement and deformation regulates the interplay between the flow induced by local isotropic deformation and the resistance to pulsation stemming from steric interaction. We reveal that this interplay leads the emergent contraction waves to spontaneously organize into a packing of pacemakers. The relaxation of these pacemakers is governed by a complex feedback between fast and slow aster defects that form in the profiles of velocity flows and of deformation gradients. By mapping contraction waves into these aster defects, we show that such waves adhere to specific constraints, including topological charge conservation and a master curve in the defect velocity statistics, that help rationalize the mechanisms underlying the relaxational dynamics.
\end{abstract}

\maketitle


{\em Introduction.}---Propagating contraction waves are observed in many biological systems, ranging from the acto-myosin cortex to some confluent tissues. For example, waves in cardiac tissues organize into various patterns during tachycardia or ventricular fibrillation~\cite{karma-annrev, Nitsan-cardiomyocyte, christoph2018electromechanical} with a specific topology~\cite{Nele2024, Nele2025}. Contraction patterns can also be found in sustained oscillations of confined epithelial cells~\cite{Peyret-oscillations-waves19}, and the collective oscillatory dynamics of electrically coupled uterine cells~\cite{Xu2015} are believed to be the basis for uterine contractions during labor~\cite{Myers2017, Maeda2013UterineCI, Elad2017}. Contractile biological systems are commonly described by active gel models~\cite{Kruse...Sekimoto2004, Joanny2009, Jonas2010, Prost2015} that combine mechanical arguments~\cite{Jonas2010, tlili2015colloquium, ishihara-pre2017,C8SM00446C, hernandez2021, grossman2022instabilities} with phenomenological theories of active matter~\cite{active-matter-review-bechinger, active-matter-marchetti, Marchetti2018, Chate2020, MIPS}. In fact, some of these models feature oscillations~\cite{martin2009pulsed, Hakim-oscillations-NatComm14, Peyret-oscillations-waves19} and wave propagation~\cite{Serra-Picamal-Natphys2012, Zaritsky-PLoS, Banerjee-Marchetti-PRL2015, tlili-RSoc18, Petrolli-PRL2019, Hino-DevCell2020, Armon-PNAS, Young1997} associated with various biological functions~\cite{kruse2011spontaneous, collinet2021programmed}.

Recently, models of deformable particles have garnered increasing attention~\cite{Peyret-oscillations-waves19, PhysRevX.6.021011, Manning-essay, PhysRevLett.129.148101, PhysRevLett.130.038202, goth2025, Zhang_2025}. For example, vertex models~\cite{farhadifar2007influence, staple2010mechanics} with mechanochemical feedback~\cite{Armon-Comm-phys, Soto2022, Hannezo-PRXLife, Shiladitya-arxiv23} have successfully described the emergence of contraction waves and pulses in confluent tissues. Dense assemblies of pulsating particles~\cite{Togashi2019, Yiwei-Etienne-PAM, Liu_2024, li2024fluidization, pineros2024biased, manacorda2023pulsating, casagrande2026a, casagrande2026b}, with sizes subject to periodic driving, also lead to contraction waves: the interplay between isotropic deformation and repulsion yields dynamical patterns reminiscent of pulsatile tissues~\cite{karma-annrev, collinet2021programmed}. Hydrodynamic studies of this pulsating active matter (PAM) have delineated field theories~\cite{Yiwei-Etienne-PAM, manacorda2023pulsating, casagrande2026a, casagrande2026b} that capture dynamical patterns distinct from those of the standard complex Ginzburg-Landau equation (CGLE)~\cite{aranson-review}.

\begin{figure}[b!]
	\includegraphics[width=\columnwidth]{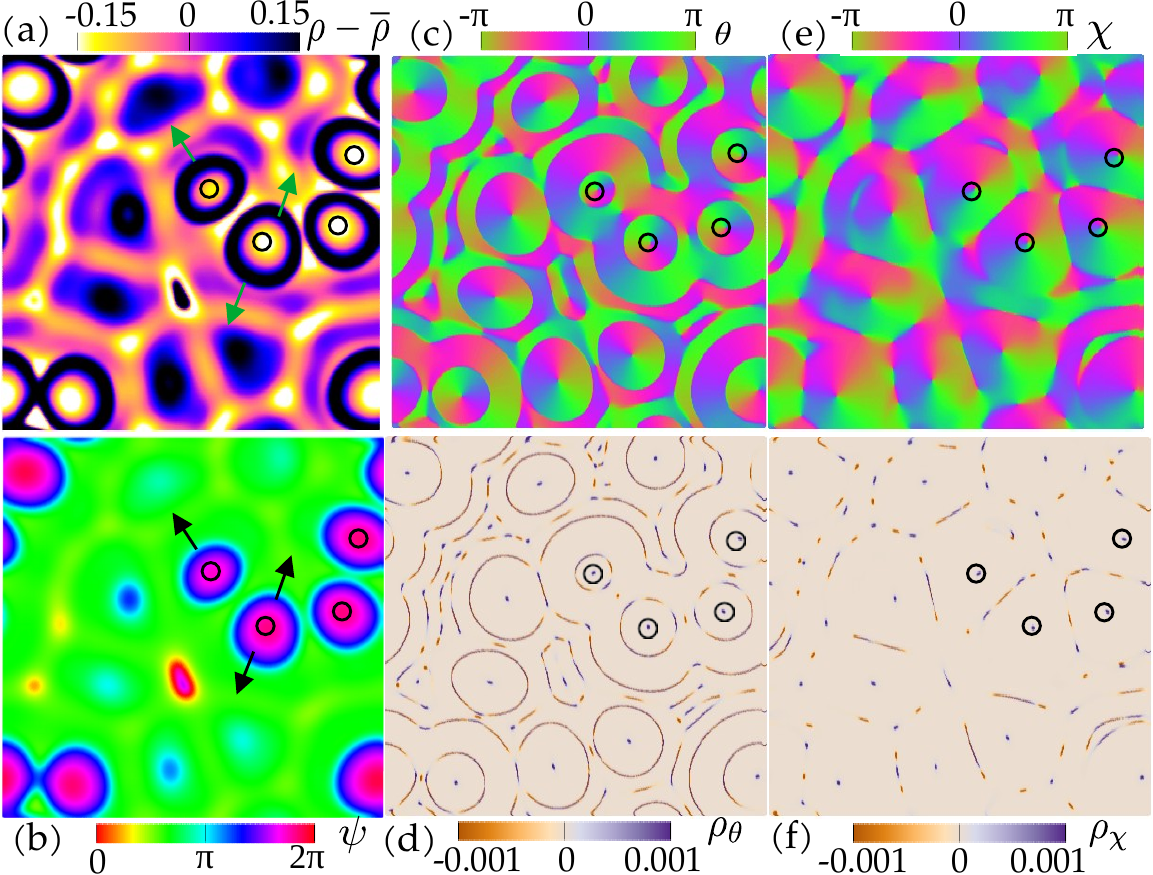}
	\caption{Contraction waves self-organize into a packing of topological defects~\cite{Movie_defects, data}: (a)~density $\rho$, (b)~phase $\psi$,  (c,f)~velocity orientation $\theta$ and topological charge density $\rho_\theta$, (e,f)~phase gradient orientation $\chi$ and topological charge density $\rho_\chi$. Markers ($\circ$) locate {\em pacemakers} where waves are triggered (as indicated by arrows) that correspond to {\em asters} in orientation profiles $(\chi,\theta)$. Parameters: $\bar\rho=1.04$, $\omega=0.125$, $L\times L=64\times64$, $dx=0.08$, $dt=0.001$.
	}
\label{fig1}
\end{figure}

In this paper, we first formulate a hydrodynamic theory that integrates the mechanochemical coupling between displacement and contraction of isotropically deformable particles. Our perspective goes beyond previous studies of PAM that neglected such a coupling~\cite{Yiwei-Etienne-PAM, manacorda2023pulsating, casagrande2026a, casagrande2026b}, and so were inadequate to describe contraction waves in density profiles. Here, we assume that pulsation is slower than stress relaxation and neighbor exchange, as reported in some confluent tissues~\cite{Zehnder2015, THIAGARAJAN2022105053}, and discard any solidification arising from high-density rigidity~\cite{Hannezo-glassy-tissue, PhysRevX.6.021011}, so the density obeys a liquid dynamics. Our coupling between density and chemical phase, which determines local isotropic deformation, captures the three main states of PAM~\cite{Togashi2019, Yiwei-Etienne-PAM, Liu_2024, li2024fluidization, pineros2024biased, manacorda2023pulsating}: (i)~a pulsating chemical phase with homogeneous density, (ii)~a constant chemical phase with homogeneous density, and (iii)~synchronized waves in density and chemical phase. The waves are periodically triggered around pacemakers that spontaneously organize into a packing configuration [Figs.~\ref{fig1}(a,b)]. We discuss in Ref.~\cite{ALP} how this specific phenomenology sets our model apart from other reaction-diffusion theories~\cite{Aranson2002, LINDNER2004321, ALIEV1996293, GANI201630, FN-review, Sakaguchi-Maeyama, tirtha-oscillator, Cross_Greenside_2009, hohenberg-review, Sakaguchi-Maeyama}.

We then examine in detail how the dynamics relaxes over many pulsations to converge towards configurations with only a few pacemakers. To this end, we demonstrate that contraction waves can be systematically associated with topological defects in the orientation profiles of velocity flows and of phase gradients [Figs.~\ref{fig1}(c,f)]. Our key insight is that the wave relaxation can then be analyzed through defect statistics. We reveal that the distribution of defect velocity obeys the same master curve even when the relaxation follows strikingly different pathways. Moreover, we unveil a coexistence between fast intermittent defects and slow stable defects whose spatial segregation regulates distinct relaxation mechanisms.

Our analysis is inspired by a body of literature that has been successful in characterizing defects in various equilibrium field theories~\cite{Mazenko1997, qi2008, Skogvoll2023, Angheluta_2021, PhysRevE.85.011153}. Here, we show how to adapt these tools to the nonequilibrium dynamics of active liquids. As such, our results provide a novel topological approach to search for generic mechanisms in the self-organization of liquids far from equilibrium.



{\em Hydrodynamic theory of pulsating liquids.}---We consider a dense assembly of pulsating particles that carry a chemical phase $\psi_i$ undergoing oscillations that can embody, for example, cellular clocks or chemical signals. Assuming that phase varies slowly between particles, we use the coarse-grained field $\psi({\bf r},t) $ to describe its large-scale behavior. We also introduce the density field $\rho({\bf r},t)$, and the velocity field ${\bf v}({\bf r},t)$ that advects density.

We neglect shear stresses~\cite{ALP}, so external forces compensate for the gradients of pressure $p$, and focus on the overdamped regime. External forces are given by friction (due to a substrate) and noise:
\begin{equation}\label{ext-force-balance}
    \partial_t \rho = -\nabla \cdot (\rho  {\bf v}) ,
    \quad
    0 = - \gamma \rho {\bf v} + \gamma \sqrt{2D_\rho} \bm\eta_\rho - \nabla p ,
\end{equation}
where $\gamma>0$, and $\bm{\eta}_\rho$ has Gaussian statistics with zero mean and unit variance: $\langle\eta_{\rho,\alpha}({\bf r},t)\eta_{\rho,\beta}({\bf r}',t') \rangle = \delta_{\alpha\beta} \delta({\bf r}-{\bf r}')\delta(t-t')$. The constitutive relation $p=p(\rho,\psi)$, which embodies the mechanochemical coupling between $\rho$ and $\psi$, follows from the free-energy density $f$:
\begin{equation}\label{eq:stress}
    \frac{p}{\rho_0} = \frac{\partial f}{\partial \rho} ,
    \quad
    f = \frac{\lambda}{2} \left( \frac{\rho -\rho_{\rm ref}}{\rho_0} \right)^2 ,
    \quad
    \frac{\rho_{\rm ref}}{\rho_0} = 1+ \epsilon \cos \psi ,
\end{equation}
where $\rho_0$ is a baseline density, and $\lambda > 0$ is the compressibility of the tissue. The density $\rho_{\rm ref}$ (at which the pressure vanishes) describes how the preferred cell area varies with the internal phase $\psi$, and $0\le\epsilon<1$ measures the strength of this modulation. In a pressure-free configuration, the density $\rho$ locally adjusts to $\rho_{\rm ref}$. In general, local deviations of $\rho$ from $\rho_{\rm ref}$ lead to pressure gradients, which in turn generate flows of $\bf v$ advecting $\rho$. This mechanism captures the displacement of particles induced by their local deformation in confluent systems~\cite{ALP}.

Individual pulsation favors oscillations of $\psi$ at the same frequency across the whole system (without imposing a uniform profile of $\psi$ a priori), and we assume that neighboring particles tend to locally synchronize their phases~\cite{Togashi2019, Yiwei-Etienne-PAM, Liu_2024, li2024fluidization, pineros2024biased, casagrande2026a, casagrande2026b, ALP}. Moreover, $\rho$ impacts $\psi$ through the mechanochemical coupling regulated by $f$. The dynamics follows as
\begin{equation}\label{eq:dyn}
\begin{aligned}
    \partial_t \rho &= \nabla \cdot \left( \frac{\rho_0}{\gamma} \nabla \frac{\partial f}{\partial \rho} + \sqrt{2 D_{\rho}} \bm {\eta}_\rho\right)  , 
    \\
    \partial_t \psi  &= \omega - \mu \frac{\partial f}{\partial \psi} + \mu\kappa \nabla^2 \psi + \sqrt{2 D_\psi} \, \eta_{\psi} , 
\end{aligned}
\end{equation}
where $\omega>0$ is the driving frequency, $\mu>0$ a kinetic coefficient, and $\kappa>0$ penalizes the formation of interfaces. The term $\partial f/\partial\rho$ comes from combining Eqs.~\eqref{ext-force-balance} and~\eqref{eq:stress}, while $\partial f/\partial\psi$ can be regarded as a density-dependent resistance to cycling. The noise $\eta_\psi$ is uncorrelated with $\bm{\eta}_\rho$, and has Gaussian statistics with zero mean and correlations given by $ \langle \eta_{\psi}({\bf r},t)  \eta_{\psi}({\bf r}',t') \rangle = \delta({\bf r}-{\bf r}')\delta(t-t')$. The dynamics follows gradient flows of $f-(\omega/\mu)\psi$ with respect to $(\rho,\psi)$ whenever $D_\psi/\mu = \gamma D_\rho/\rho_0$.

The equilibrium limit ($\omega=0$ and $D_\psi/\mu = \gamma D_\rho/\rho_0$) reduces to a version of model C~\cite{hohenberg-review, Bray1994, Cross_Greenside_2009} that neither accommodates any instability nor steady currents. In fact, our theory is not the most general nonequilibrium extension of model C; for instance, see Ref.~\cite{Maryshev-actmodelC} for an active coupling between density and nematic fields, Ref.~\cite{Erwin-actmodelC} for active emulsions, and Refs.~\cite{mmz3-kbrv, cocconi2025} for models of mechanochemical feedback where particle configurations affect their activity.

For simplicity, we focus here on noiseless dynamics ($D_\psi=0=D_\rho$), and take $(\rho_0,\gamma,\mu,\lambda,\kappa,2\epsilon)$ all equal to $1$; see \cite{ALP} for discussion on non-dimensionalized dynamics. The dynamics is then only controlled by the total density $\bar\rho=\frac 1 V \int_V \rho({\bf r},t) d{\bf r}$ and the driving $\omega$. Our model contains three states that are the key features of PAM~\cite{Yiwei-Etienne-PAM, manacorda2023pulsating, Liu_2024, li2024fluidization, pineros2024biased, casagrande2026a, casagrande2026b}: homogeneous arrest at large $(\omega^{-1},|\bar\rho-1|)$, homogeneous cycles at small $(\omega^{-1},|\bar\rho-1|)$, and propagating waves in intermediate regimes [Fig.~\ref{fig1}]. The detailed phase diagram, including analytical prediction of phase boundaries, is presented in Ref.~\cite{ALP}.

In what follows, we focus on the regime of waves. For simulations, we use a finite-difference scheme in two dimensions~\cite{thampi-ronojoy} that conforms to the conservation law and respects isotropy of the underlying lattice, see Ref.~\cite{ALP} for details. Our aim is to uncover how topological defects in the orientation profile of specific fields regulate the complex self-organization of waves. 



{\em Topological defects in orientation profiles.}---Our contraction waves are not associated with any net mass transport. Introducing the orientation $\theta$ of the velocity field $\bf v$ that advects density $\rho$ [Eq.~\eqref{ext-force-balance}]:
\begin{equation}\label{eq:theta}
    \tan\theta = {\rm v}_y/{\rm v}_x ,
\end{equation}
we observe that waves drive velocity flows that periodically change directions: as a wave is triggered and propagates radially, it reverses the orientation of $\bf v$ (namely, $\theta$ shifts by $\pi$), so the density $\rho$ gets advected in the opposite direction; see~Fig.~\ref{fig1}(c), and videos in Ref.~\cite{Movie_defects}. These waves form {\em asters} in the orientation profile $\theta$, namely polar defects with charge $+1$, that are centered around pacemakers of propagating waves. In fact, similar asters also appear in the orientation profile $\chi$ of the gradient of the chemical field $\psi$ [Fig.~\ref{fig1}(e)]:
\begin{equation}\label{eq:chi}
	\tan\chi = (\nabla_y \psi)/(\nabla_x \psi) .
\end{equation}
The total charge of defects in the orientation profiles $(\theta,\chi)$ must vanish for periodic boundary conditions in two dimensions [Appendix~A], so the patterns in $(\theta,\chi)$ entail a spatial coexistence between $+1$ and $-1$ charged defects.

To examine defect statistics, we consider the topological charge density fields $\rho_{\rm X}$ associated with each one of the orientations ${\rm X} \in \{\theta,\chi\}$: 
\begin{equation}
	\rho_{\rm X} = (\varepsilon_{\alpha\beta}/\pi)(\partial_\alpha \cos{\rm X})(\partial_\beta\sin{\rm X}) ,
\end{equation}
where $\varepsilon_{\alpha\beta}$ is the Levi-Civita tensor, and we assume an implicit summation over the Cartesian coordinates $(\alpha, \beta)$. Integrating $\rho_{\rm X}$ over the arbitrary surface $V_{\rm d}$ that contains a defect with charge $q_{\rm X}$ yields
\begin{equation}
	\int_{V_{\rm d}} \rho_{\rm X} d{\bf r} = \frac{1}{2\pi}\oint_{\partial V_{\rm d}} d{\rm X} = q_{\rm X} ,
\end{equation}
where $\partial V_{\rm d}$ is the line enclosing $V_{\rm d}$~\cite{Mazenko1997, qi2008, Skogvoll2023, PhysRevE.85.011153, Angheluta_2021}. The density $\rho_{\rm X}$ vanishes wherever the orientation profile ${\rm X}$ is smooth, while $\pm 1$ defects yield non-zero values of $\rho_{\rm X}$, where $-1$ defects correspond to saddles of the corresponding orientation $\rm X$. In practice, $\rho_\chi$ only detects some isolated defects, whereas $\rho_\theta$ features additional line defects along the wavefronts where $\bf v$ vanishes [Figs.~\ref{fig1}(c-f)].

Given the definition ${\bf v} = (1/\gamma)\nabla p$ in terms of the pressure $p$ [Eq.~\eqref{ext-force-balance}], it follows that both $(\theta,\chi)$ are orientations of gradient vectors. This property enforces a series of topological constraints given by Morse theory~\cite{morse}. Vortex and spiral defects are excluded, so that $q_{\rm X}=+1$ defects can only form into asters [Appendix~A]; in contrast, defects of arbitrary polar vectors can be either asters, vortices, or spirals~\cite{Chaikin, active-matter-marchetti, Angheluta2026}. Moreover, for periodic boundary conditions in two dimensions, the steady state associated with a single pacemaker features (at least) two pairs of $\pm 1$ defects, namely four defects in total [Appendix~A]; in contrast, the ground state of standard spin models and their polar/nematic hydrodynamic theories can be fully ordered without any defects~\cite{Chaikin, Bray1994}.

\begin{figure}
	\includegraphics[width=\columnwidth]{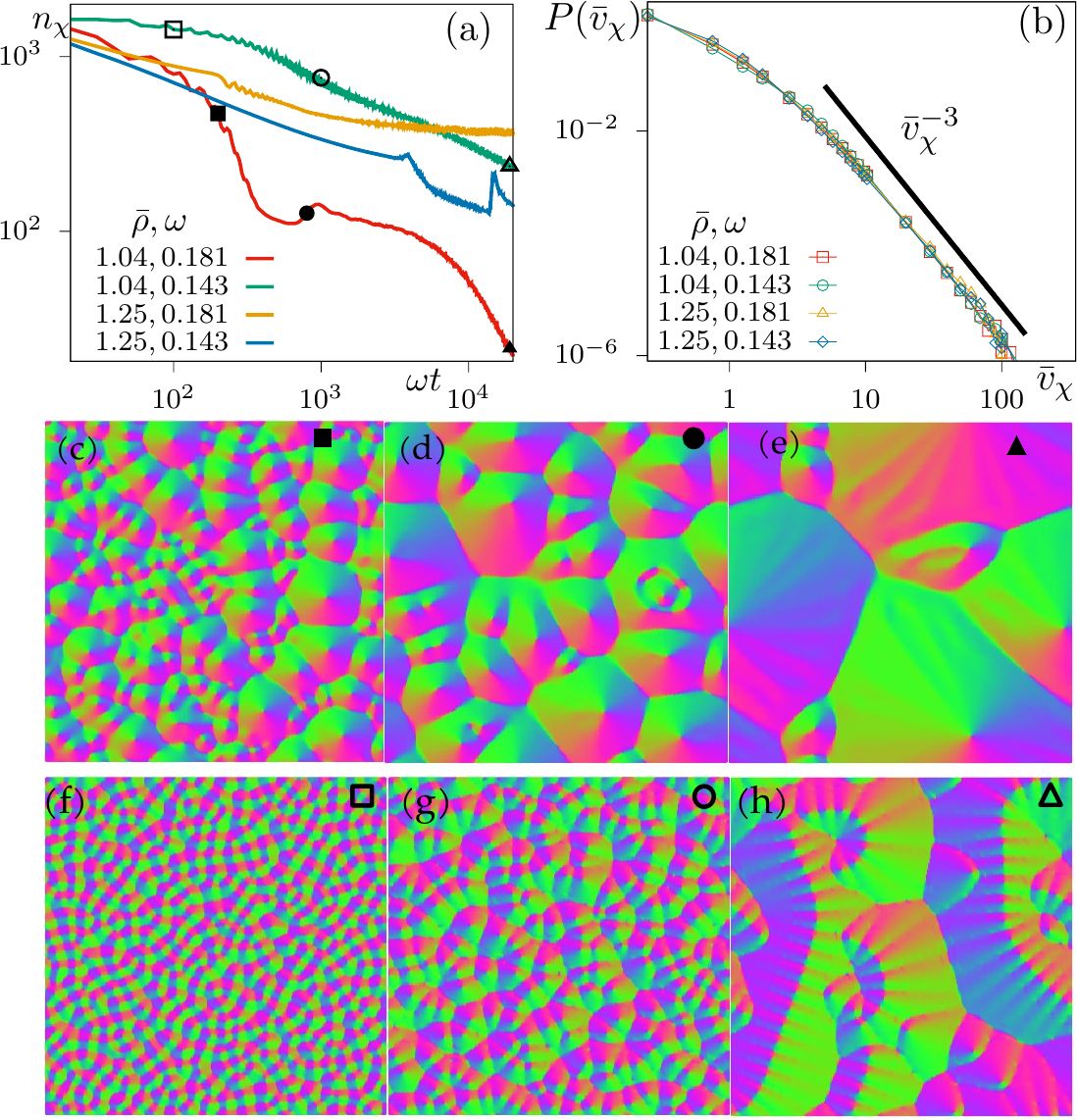}
	\caption{(a)~Trajectories of number of defects $n_\chi$ for various $(\bar \rho, \omega)$, averaged over $128$ random-phase initial distributiondistributionrealizations. Markers refer to snapshots in panels~(c-h).
	(b)~Probability distribution $P(\bar v_{\chi})$ of scaled defect velocity $\bar v_{\chi}$ follows a master curve.
	(c-h)~Snapshots of phase gradient orientation $\chi$~\cite{Movie_defects}. Same colorbar as in Fig.~\ref{fig1}.
	Parameters: $L\times L=256\times 256$, $dx=0.8$, $dt=0.01$.
	}
	\label{fig2}
\end{figure}

The conservation of topological charge ($\int_V \rho_{\rm X} d{\bf r}=0$) entails that $\rho_{\rm X}$ obeys a conservation law:
\begin{equation}\label{eq:vd}
	\partial_t \rho_{\rm X} = - \nabla\cdot(\rho_{\rm X}{\bm v}_{\rm X}) ,
\end{equation}
where the topological velocity field ${\bm v}_{\rm X}$ advecting defects read~\cite{Mazenko1997, qi2008, Skogvoll2023}
\begin{equation}\label{eq:vel}
\begin{aligned}
   {v}_{{\rm X},x} &= \frac{1}{\pi \rho_{\rm X}} \Big[ (\partial_t \sin {\rm X}) (\partial_y \cos {\rm X}) - (\partial_t \cos {\rm X}) (\partial_y \sin {\rm X}) \Big] ,
   \\
   {v}_{{\rm X},y} &= \frac{1}{\pi \rho_{\rm X}} \Big[ (\partial_t \cos {\rm X}) (\partial_x \sin {\rm X}) - (\partial_t \sin {\rm X})(\partial_x \cos {\rm X}) \Big] . 
\end{aligned}
\end{equation}
The number of defects $n_{\rm X}$, defined by
\begin{equation}\label{eq:number}
    n_{\rm X}(t) = \frac 1 2 \int_{V} |\rho_{\rm X}({\bf r},t)| d{\bf r} ,
\end{equation}
is not conserved, so trajectories of $n_{\rm X}(t)$ characterize the relaxational dynamics [Fig.~\ref{fig2}(a)]. These topological definitions [Eqs.~(\ref{eq:theta}-\ref{eq:number})] can be extended to other boundary conditions and higher spatial dimensions.

In what follows, we show that a proper analysis of defect statistics helps unveil the mechanisms at play in the relaxation of the domains centered around pacemakers. We focus on the defect in orientation profile $\chi$, since the corresponding charge density $\rho_\chi$ does not feature any defect lines along contraction wavefronts.

{\em Defects regulate coarsening of pacemaker domains.}---Starting from a disordered phase $\psi$ and homogeneous density $\rho$, the system spontaneously forms aster defects in the orientation profiles $(\theta,\chi)$ [Figs.~\ref{fig1}(c-f)]. We now examine how, for parameter values $(\bar\rho,\omega)$ that feature wave formation, the relaxation of aster domains is regulated by the propagation of waves.

Periodic collisions of waves lead asters to arrange into specific packing configurations. The typical distance between pacemakers defines the size of domains containing a single aster. These pacemaker domains relax much slower than the periodic wave collisions. We characterize this relaxation in terms of the time evolution of the number of defects $n_\chi$ [Eq.~\eqref{eq:number}]: it features high frequency oscillations stemming from periodic wave collisions, and an overall decay at large times [Fig.~\ref{fig2}(a)].

For all values of $(\bar\rho,\omega)$, the decrease of $n_\chi(t)$ shows that the number of asters reduces, and correspondingly the pacemaker domains coarsen in time. For each value $(\bar \rho, \omega)$, the decay profile $n_\chi(t)$ is robust over realizations, showing that the dynamics follows a specific relaxation pathway. These pathways strikingly differ with $(\bar\rho,\omega)$: for some values, $n_\chi(t)$ saturates to a high plateau value; for others, we report a non-monotonic behavior of $n_\chi(t)$ that correspond to changes in dynamical state [Appendix~B]. Moreover, the transition between waves and arrested state is associated with a specific power-law $n_\chi(t)\sim 1/\sqrt{t}$ [Appendix~B].

The large variety of relaxation pathways illustrates the complex behavior of the emergent defect dynamics. To further characterize defect statistics, we evaluate defect speed: 
\begin{equation}
    \bar v_{\chi}({\bf r},t) = \frac{|{\bm v}_{\chi} ({\bf r},t)|}{\langle v_{\chi} \rangle_t} ,
    \quad
    \langle v_{\chi} \rangle_t = \int_V |{\bm v}_{\chi} ({\bf r},t) | \frac{d{\bf r}}{V} ,
\end{equation}
where the spatially averaged value $\langle v_{\rm X} \rangle_t$ is evaluated at a given instant $t$. For all values of $(\bar\rho,\omega)$, the distribution $P(\bar v_\chi)$ collapses onto a master curve whose tail is consistent with a power-law decay: $P(\bar v_\chi)\sim \bar v_\chi^{-3}$ [Fig.~\ref{fig2}(b)]. This master curve holds even when relaxation pathways are strikingly different; for instance, when asters organize either into an almost regular packing configuration [Figs.~\ref{fig2}(c-e)], or into aligned defects [Figs.~\ref{fig2}(f-h)].

Defect-speed distributions with a similar power-law tail have been reported in various systems, ranging from equilibrium (e.g., Ginzburg-Landau field theories and XY spin models~\cite{Mazenko1997, PhysRevE.85.011153}) to nonequilibrium models (e.g., complex Ginzburg-Landau field theories and experiments on living-cell membranes~\cite{Tan2020}): the specific power-law exponent has been related to defect annihilation and symmetries of the underlying dynamics~\cite{Bray1994}. Although our waves differ from these other patterns, the similarity in velocity distribution tails suggests that our defects and theirs share analogous annihilation mechanisms.

In short, although the coarsening of pacemaker domains can follow various dynamical pathways, the defect velocity statistics always obeys the same master curve. This result suggests that similar defect interaction mechanisms are at play in each pathway. To identify such mechanisms, we now isolate and analyze the topology of a single pacemaker, which embodies the simplest configuration shared by all pathways.


\begin{figure}
    \includegraphics[width=\columnwidth]{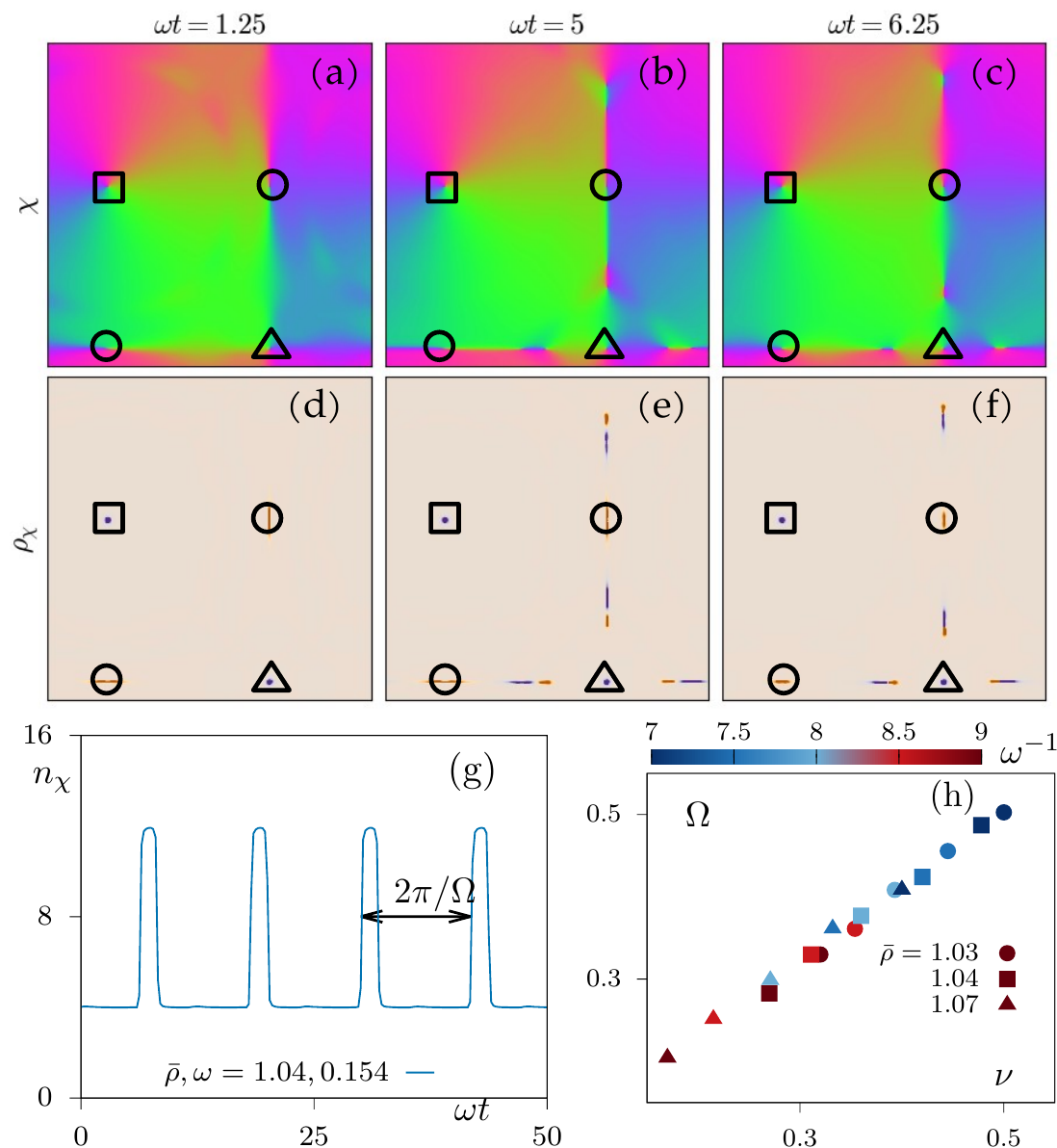}
    \caption{Snapshots of (a-c)~orientation field $\chi$ and (d-f)~defect charge density $\rho_\chi$ in a steady-state configuration with a single pacemaker~\cite{Movie_defects}. Markers refer to wave source ($\square$, $q_\chi=+1$), wave sink ($\triangle$, $q_\chi=+1$), and boundaries of pacemaker domain ($\bigcirc$, $q_\chi=-1$). Same colorbars as in Fig.~\ref{fig1}.
    (g)~Trajectories of number of defects $n_\chi(t)$ oscillate at frequency $\Omega$. Parameters: $\bar\rho=1.04$, $\omega^{-1}=6.5$, $L\times L=32\times 32$, $dx=0.08$, $dt=0.001$.
    (h)~Frequency $\Omega$ and phase current $\nu$ for various $(\bar \rho, \omega)$.
    }
\label{fig3}
\end{figure}

{\em Coexistence between fast and slow defects.}---We consider steady-state configurations with a single pacemaker, and examine how defects interact [Figs.~\ref{fig3}(a-f)]. The number of defects oscillates from $n_\chi(t)=4$ (i.e., minimum number of defects for curl-free polar vectors [Appendix~A]) to $n_\chi(t)>4$ [Fig.~\ref{fig3}(g)]. We distinguish stable defects that remain present at all times from intermittent defects that are periodically created and annihilated. When $n_\chi(t)=4$, only stable defects are present [Figs.~\ref{fig3}(a,d)]. They organize into separated locations: wave source ($q_\chi=+1$), wave sink ($q_\chi=+1$), and boundaries of pacemaker domains ($q_\chi=-1$); the charge is identical for source and sink but they can be differentiated through $\nabla^2 \psi$ as they represent, respectively, local maxima and minima of $\psi$. Waves periodically collide at boundaries of pacemaker domains, which in turn yields the formation of several pairs of intermittent defects with opposite charges $q_\chi=\pm1$ [Figs.~\ref{fig3}(b,e)]. Stable defects barely move, while intermittent defects rapidly move along the pacemaker domain boundaries towards the sink of waves.

The periodic creation and annihilation of intermittent defects sets the oscillation of $n_\chi(t)$ [Fig.~\ref{fig3}(g)]. We extract the oscillation frequency $\Omega$ as the dominant mode in the spectral power spectrum of $n_\chi(t)$. In fact, $n_\theta(t)$ oscillates with the same frequency $\Omega$. We report that $\Omega$ correlates with the phase current
\begin{equation}\label{nu}
    \nu = \int_V \frac{\langle \partial_t \psi \big\rangle}{\omega} \frac{d{\bf r}}{V}
\end{equation}
for all values of $(\bar\rho,\omega)$ [Fig.~\ref{fig3}(h)], showing that oscillations of $n_\chi(t)$, also present in configurations with many defects [Fig.~\ref{fig2}(a)], are controlled by the speed at which the phase $\psi$ cycles. This result supports that creation, annihilation, and displacement of intermittent defects is regulated by periodic collisions of waves.

In short, slow stable defects (that organize into a packing configuration) are spatially segregated from fast intermittent defects  (that periodically nucleate and annihilate at aster boundaries). While aster packing configurations are reminiscent of active foams reported in constant-density flocks~\cite{Toner2012, Besse2022} and non-reciprocal XY models~\cite{Dadhichi2020, dopierala2025, maitra2025, popli2025}, the main novelty here stems from the coexistence of these slow asters with other fast defects controlled by wave collisions.




{\em Discussion.}---Our hydrodynamics describes the large-scale behavior of active deformable particles~\cite{ALP} whose pulsation can stem from either an explicit microscopic drive~\cite{Yiwei-Etienne-PAM, Liu_2024, li2024fluidization, pineros2024biased, manacorda2023pulsating, casagrande2026a, casagrande2026b} or some emergent internal cycles~\cite{Hannezo-PRXLife, Shiladitya-arxiv23}. Other models coupling complex chemical and density fields could be considered. In fact, the definition of topological charge density~\cite{Mazenko1997, qi2008, Skogvoll2023} is agnostic to model details, so it can be straightforwardly deployed in experimental systems, such as biological tissues~\cite{karma-annrev, collinet2021programmed} where velocity and contraction flows are experimentally accessible~\cite{SUTHERLAND_99, THIAGARAJAN2022105053, tang2025}. Therefore, we anticipate that our approach will trigger further topological analysis in active liquids, inspired, for instance, by recent experiments~\cite{tang2025}.

Our topological analysis reveals constraints on defect dynamics: Morse theory only allows for asters (i.e., sources and sinks of waves) and $-1$ defects (i.e., saddles of orientations); charge conservation enforces that fast intermittent defects are created by pairs; defect velocity statistics follows a master curve. These results provide key insights for future modelling of defect dynamics in active liquids, in line with previous works on defects in active nematics~\cite{Shankar2018, Shankar2019, Shankar22}. As such, our study contributes to a long-term agenda for devising strategies that control nonequilibrium liquids through defects, using, for instance, recent tools of optimal control theory~\cite{Norton2020, Shankar2022, Davis2025, soriani2025, krishnan2024, fbgp-qpvv, alvarado2025, D4SM00547C} and inspired by results on active turbulence~\cite{active-turbulence-alert, active-turbulence-yeomans, Shankar2024, radhakrishnan2025}.

\acknowledgments{We acknowledge insightful discussion with Luiza Angheluta. \'E.F., T.D., and T.B. are supported through the Luxembourg National Research Fund (FNR), grant references 14389168 and C22/MS/17186249. L.C. acknowledges funding from the Ernest Oppenheimer Fund. We acknowledge support from grant NSF PHY-2309135 to the Kavli Institute for Theoretical Physics (KITP).}


{\em Appendix~A:~Topology of curl-free polar vectors.}---The topology of defects in the profile of a polar vector $\bf p$ can be characterized by examining weak deviations $\bm\xi$ around the defect core ${\bf r}_{\rm d}$:
\begin{equation}\label{eq:p}
	p_\alpha({\bf r}_{\rm d} + {\bm\xi}) = {\bf\xi}_\beta \, (\nabla_{\beta} p_\alpha) ({\bf r}_{\rm d}) + {\cal O}({\bf\xi}^2) ,
\end{equation}
where we have used $p_\alpha({\bf r}_{\rm d})=0$ (since polarity ${\bf p}$ vanishes at defect locations), and we assume implicit summation over repeated indices. For defect charge $+1$, the eigenvalues of the matrix $(\nabla_{\beta} p_\alpha) ({\bf r}_{\rm d})$ determine the nature of the defect: either aster, spiral, or vortex~\cite{Chaikin}. We express the curl-free polar vector $\bar{\bf p}= \nabla\phi$ in terms of the scalar field $\phi$ so that $\nabla\times\bar{\bf p}=\bm 0$, from which we deduce that the expansion in Eq.~\eqref{eq:p} can now be written with the Hessian matrix ${\mathbb H}$ as 
\begin{equation}
	\bar p_\alpha({\bf r}_{\rm d} + {\bm\xi}) = {\bf\xi}_\beta \, {\mathbb H}_{\alpha\beta} ({\bf r}_{\rm d}) + {\cal O}({\bf\xi}^2) , 
	\quad
	{\mathbb H}_{\alpha\beta} = \nabla_\alpha \nabla_\beta \phi .
\end{equation}
By construction, $\mathbb H$ is symmetric with real eigenvalues and orthogonal eigenvectors: this property precludes any vortex and spiral in the defect profile of $\bar{\bf p}$. In fact, minima and maxima of $\phi$ correspond to asters with charge $+1$, whereas saddles of $\phi$ correspond to $-1$ defects.

Morse inequalities provide further constraints on the number of defects depending on the topology of the underlying space; in fact, the number of minima, maxima, and saddles of $\phi$ are bounded by Betti numbers~\cite{morse}. For a torus (i.e., two-dimensional space with periodic boundary conditions), these bounds enforce at least one minimum, one maximum, and two saddles; in contrast, for a sphere, one gets at least one minimum and one maximum.


\begin{figure}
	\includegraphics[width=0.99\columnwidth]{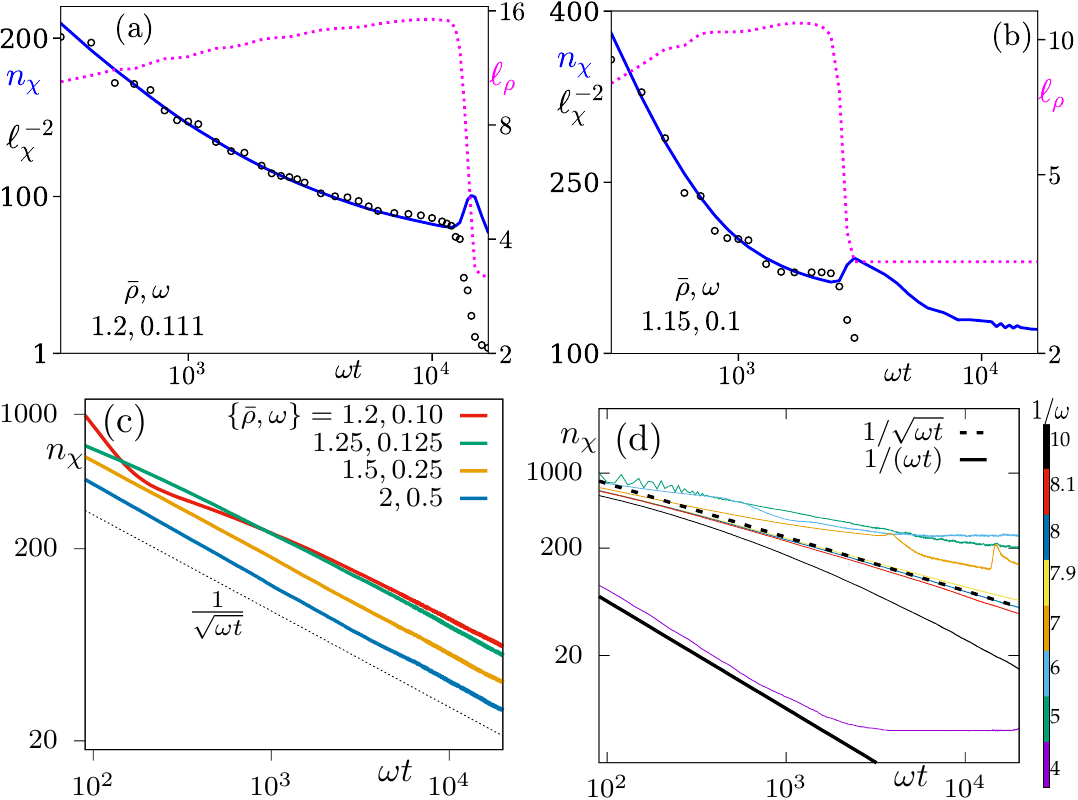}
	\caption{(a-b)~Trajectories of number of defects $n_\chi(t)$ (solid line), correlation lengths for density $\ell_\rho$ (dotted line) and for orientation $\ell_\chi$ (circular markers). Data averaged over $128$ random-phase initial realizations.
	(c)~For wave-to-arrest transition $\omega=\mu\lambda\epsilon|\bar \rho -1|$, the relaxation obeys $n_\chi(t) \sim 1/\sqrt{t}$.
	(d)~Relaxation laws $n_\chi(t)$ for stable cycles $\omega^{-1}=4$, stable arrest $\omega^{-1}=10$, stable waves $\omega^{-1}=(5,6,7)$, and close to arrest-to-wave transition $\omega^{-1}=(7.9,8,8.1)$.
	Parameters: $\bar\rho=1.25$, $L\times L=256\times 256$, $dx=0.8$, $dt=0.01$.
	}
	\label{fig4}
\end{figure}

{\em Appendix~B:~Defect dynamics and coarsening.}---The time evolution of the number of defects $n_{\chi}(t)$ can feature non-monotonic behavior for some values of $(\bar\rho,\omega)$ [Fig.~\ref{fig2}(a)]. To relate these features with changes in dynamical states, we examine the relaxation of correlation lengths $(\ell_\rho, \ell_\chi)$ of density and orientation, respectively. We measure these lengths as the first zero of the corresponding correlation functions:
\begin{equation}\label{corr_func_defn}
\begin{aligned}
	C_\rho(r) &= \big\langle \big[ \rho({\bf x}+{\bf r}, t)-\bar\rho\big]\big[\rho({\bf x}, t)-\bar\rho\big] \big\rangle,
	\\  
	C_{\chi}(r) &= \big\langle \big[{\bf e}({\bf x}+{\bf r}, t)-\bar{\bf e}(t)\big]\cdot\big[ {\bf e}({\bf x},t)-\bar{\bf e}(t) \big] \big\rangle \ ,
\end{aligned}
\end{equation}
where ${\bf e} = (\cos \chi, \sin \chi) = \nabla\psi / |\nabla\psi|$, and
\begin{equation}
	\bar\rho = \int_V \rho({\bf r}, t) \frac{d{\bf r}}{V} ,
	\quad
	\bar{\bf e}(t) = \int_V {\bf e}({\bf r}, t) \frac{d{\bf r}}{V} .
\end{equation}
We now discuss how the trajectories of $(n_\chi, \ell_\rho, \ell_\chi)$ help disentangle the various stages of relaxation towards a steady-state configuration.

Before waves spontaneously form, the relaxation corresponds to the formation of density and orientation domains following a spinodal instability~\cite{ALP}: the density correlation length $l_\rho(t)$ increases, while the number of defects $n_\chi(t)$ decreases [Figs.~\ref{fig4}(a-b)]. Moreover, $n_\chi(t)$ scales like $\ell_\chi^{-2}(t)$, showing that this initial coarsening of domains is controlled by the kinetics of defect annihilation, as in standard scenarios for relaxation of passive defects~\cite{Bray1994}. The emergence of waves yields a clear signature in the relaxation: $(\ell_\rho, \ell_\chi^{-2})$ decrease sharply, while $n_\chi$ increases at the same instant. In this regime, waves trigger the nucleation of defects, so the relaxation kinetics is no longer controlled only by defect annihilation.

As $\bar\rho$ increases and $\omega$ decreases, the waves slow down until the system undergoes a transition to an arrested state when $\omega = \mu \lambda \epsilon|\bar \rho -1|$~\cite{ALP}. For the corresponding values of $(\bar\rho,\omega)$, we measure a specific relaxation law: $n_\chi(t) \sim 1/\sqrt{t}$ [Fig.~\ref{fig4}(c)]. Varying $\omega$ at fixed $\bar\rho=1.25$ [Fig.~\ref{fig4}(d)], we report how the relaxation law changes for $n_\chi(t)$. Deep within regions of stable arrest and stable cycles, where the system stays homogeneous in steady state, the relaxation obeys $n_\chi(t) \sim 1/t$. Close to arrest-to-wave transition, there is a smooth transition to the law $n_\chi(t)\sim 1/\sqrt{t}$.

\bibliography{references-dpm}

\begin{thebibliography}{110}%
\makeatletter
\providecommand \@ifxundefined [1]{%
 \@ifx{#1\undefined}
}%
\providecommand \@ifnum [1]{%
 \ifnum #1\expandafter \@firstoftwo
 \else \expandafter \@secondoftwo
 \fi
}%
\providecommand \@ifx [1]{%
 \ifx #1\expandafter \@firstoftwo
 \else \expandafter \@secondoftwo
 \fi
}%
\providecommand \natexlab [1]{#1}%
\providecommand \enquote  [1]{``#1''}%
\providecommand \bibnamefont  [1]{#1}%
\providecommand \bibfnamefont [1]{#1}%
\providecommand \citenamefont [1]{#1}%
\providecommand \href@noop [0]{\@secondoftwo}%
\providecommand \href [0]{\begingroup \@sanitize@url \@href}%
\providecommand \@href[1]{\@@startlink{#1}\@@href}%
\providecommand \@@href[1]{\endgroup#1\@@endlink}%
\providecommand \@sanitize@url [0]{\catcode `\\12\catcode `\$12\catcode
  `\&12\catcode `\#12\catcode `\^12\catcode `\_12\catcode `\%12\relax}%
\providecommand \@@startlink[1]{}%
\providecommand \@@endlink[0]{}%
\providecommand \url  [0]{\begingroup\@sanitize@url \@url }%
\providecommand \@url [1]{\endgroup\@href {#1}{\urlprefix }}%
\providecommand \urlprefix  [0]{URL }%
\providecommand \Eprint [0]{\href }%
\providecommand \doibase [0]{http://dx.doi.org/}%
\providecommand \selectlanguage [0]{\@gobble}%
\providecommand \bibinfo  [0]{\@secondoftwo}%
\providecommand \bibfield  [0]{\@secondoftwo}%
\providecommand \translation [1]{[#1]}%
\providecommand \BibitemOpen [0]{}%
\providecommand \bibitemStop [0]{}%
\providecommand \bibitemNoStop [0]{.\EOS\space}%
\providecommand \EOS [0]{\spacefactor3000\relax}%
\providecommand \BibitemShut  [1]{\csname bibitem#1\endcsname}%
\let\auto@bib@innerbib\@empty
\bibitem [{\citenamefont {Karma}(2013)}]{karma-annrev}%
  \BibitemOpen
  \bibfield  {author} {\bibinfo {author} {\bibfnamefont {Alain}\ \bibnamefont
  {Karma}},\ }\bibfield  {title} {\enquote {\bibinfo {title} {Physics of
  cardiac arrhythmogenesis},}\ }\href {\doibase
  10.1146/annurev-conmatphys-020911-125112} {\bibfield  {journal} {\bibinfo
  {journal} {Annu. Rev. Condens. Matter Phys.}\ }\textbf {\bibinfo {volume}
  {4}},\ \bibinfo {pages} {313--337} (\bibinfo {year} {2013})}\BibitemShut
  {NoStop}%
\bibitem [{\citenamefont {Nitsan}\ \emph {et~al.}(2016)\citenamefont {Nitsan},
  \citenamefont {Drori}, \citenamefont {Lewis}, \citenamefont {Cohen},\ and\
  \citenamefont {Tzlil}}]{Nitsan-cardiomyocyte}%
  \BibitemOpen
  \bibfield  {author} {\bibinfo {author} {\bibfnamefont {Ido}\ \bibnamefont
  {Nitsan}}, \bibinfo {author} {\bibfnamefont {Stavit}\ \bibnamefont {Drori}},
  \bibinfo {author} {\bibfnamefont {Yair~E.}\ \bibnamefont {Lewis}}, \bibinfo
  {author} {\bibfnamefont {Shlomi}\ \bibnamefont {Cohen}}, \ and\ \bibinfo
  {author} {\bibfnamefont {Shelly}\ \bibnamefont {Tzlil}},\ }\bibfield  {title}
  {\enquote {\bibinfo {title} {Mechanical communication in cardiac cell
  synchronized beating},}\ }\href {\doibase 10.1038/nphys3619} {\bibfield
  {journal} {\bibinfo  {journal} {Nat. Phys.}\ }\textbf {\bibinfo {volume}
  {12}},\ \bibinfo {pages} {472--477} (\bibinfo {year} {2016})}\BibitemShut
  {NoStop}%
\bibitem [{\citenamefont {Christoph}\ \emph {et~al.}(2018)\citenamefont
  {Christoph}, \citenamefont {Chebbok}, \citenamefont {Richter}, \citenamefont
  {Schr{\"o}der-Schetelig}, \citenamefont {Bittihn}, \citenamefont {Stein},
  \citenamefont {Uzelac}, \citenamefont {Fenton}, \citenamefont {Hasenfu{\ss}},
  \citenamefont {Gilmour~Jr} \emph {et~al.}}]{christoph2018electromechanical}%
  \BibitemOpen
  \bibfield  {author} {\bibinfo {author} {\bibfnamefont {Jan}\ \bibnamefont
  {Christoph}}, \bibinfo {author} {\bibfnamefont {Mohammed}\ \bibnamefont
  {Chebbok}}, \bibinfo {author} {\bibfnamefont {Claudia}\ \bibnamefont
  {Richter}}, \bibinfo {author} {\bibfnamefont {Johannes}\ \bibnamefont
  {Schr{\"o}der-Schetelig}}, \bibinfo {author} {\bibfnamefont {Philip}\
  \bibnamefont {Bittihn}}, \bibinfo {author} {\bibfnamefont {Sebastian}\
  \bibnamefont {Stein}}, \bibinfo {author} {\bibfnamefont {Ilja}\ \bibnamefont
  {Uzelac}}, \bibinfo {author} {\bibfnamefont {Flavio~H}\ \bibnamefont
  {Fenton}}, \bibinfo {author} {\bibfnamefont {Gerd}\ \bibnamefont
  {Hasenfu{\ss}}}, \bibinfo {author} {\bibfnamefont {RF}~\bibnamefont
  {Gilmour~Jr}},  \emph {et~al.},\ }\bibfield  {title} {\enquote {\bibinfo
  {title} {Electromechanical vortex filaments during cardiac fibrillation},}\
  }\href {\doibase 10.1038/nature26001} {\bibfield  {journal} {\bibinfo
  {journal} {Nature}\ }\textbf {\bibinfo {volume} {555}},\ \bibinfo {pages}
  {667--672} (\bibinfo {year} {2018})}\BibitemShut {NoStop}%
\bibitem [{\citenamefont {Duytschaever}\ \emph {et~al.}(2024)\citenamefont
  {Duytschaever}, \citenamefont {den Abeele}, \citenamefont {Carlier},
  \citenamefont {Bezerra}, \citenamefont {Verstraeten}, \citenamefont
  {Lootens}, \citenamefont {Desplenter}, \citenamefont {Okenov}, \citenamefont
  {Nezlobinsky}, \citenamefont {Shah}, \citenamefont {Haas}, \citenamefont
  {Luik}, \citenamefont {Martens}, \citenamefont {Haddad}, \citenamefont
  {Smet}, \citenamefont {Becker}, \citenamefont {Francois}, \citenamefont
  {de~Waroux}, \citenamefont {Tavernier}, \citenamefont {Knecht}, \citenamefont
  {Hendrickx},\ and\ \citenamefont {Vandersickel}}]{Nele2024}%
  \BibitemOpen
  \bibfield  {author} {\bibinfo {author} {\bibfnamefont {Mattias}\ \bibnamefont
  {Duytschaever}}, \bibinfo {author} {\bibfnamefont {Robin~Van}\ \bibnamefont
  {den Abeele}}, \bibinfo {author} {\bibfnamefont {Niels}\ \bibnamefont
  {Carlier}}, \bibinfo {author} {\bibfnamefont {Arthur~Santos}\ \bibnamefont
  {Bezerra}}, \bibinfo {author} {\bibfnamefont {Bjorn}\ \bibnamefont
  {Verstraeten}}, \bibinfo {author} {\bibfnamefont {Sebastiaan}\ \bibnamefont
  {Lootens}}, \bibinfo {author} {\bibfnamefont {Karel}\ \bibnamefont
  {Desplenter}}, \bibinfo {author} {\bibfnamefont {Arstanbek}\ \bibnamefont
  {Okenov}}, \bibinfo {author} {\bibfnamefont {Timur}\ \bibnamefont
  {Nezlobinsky}}, \bibinfo {author} {\bibfnamefont {Dipen}\ \bibnamefont
  {Shah}}, \bibinfo {author} {\bibfnamefont {Annika}\ \bibnamefont {Haas}},
  \bibinfo {author} {\bibfnamefont {Armin}\ \bibnamefont {Luik}}, \bibinfo
  {author} {\bibfnamefont {Jordi}\ \bibnamefont {Martens}}, \bibinfo {author}
  {\bibfnamefont {Milad~El}\ \bibnamefont {Haddad}}, \bibinfo {author}
  {\bibfnamefont {Maarten~De}\ \bibnamefont {Smet}}, \bibinfo {author}
  {\bibfnamefont {Benjamin~De}\ \bibnamefont {Becker}}, \bibinfo {author}
  {\bibfnamefont {Clara}\ \bibnamefont {Francois}}, \bibinfo {author}
  {\bibfnamefont {Jean-Benoit Le~Polain}\ \bibnamefont {de~Waroux}}, \bibinfo
  {author} {\bibfnamefont {Rene}\ \bibnamefont {Tavernier}}, \bibinfo {author}
  {\bibfnamefont {Sebastien}\ \bibnamefont {Knecht}}, \bibinfo {author}
  {\bibfnamefont {Sander}\ \bibnamefont {Hendrickx}}, \ and\ \bibinfo {author}
  {\bibfnamefont {Nele}\ \bibnamefont {Vandersickel}},\ }\bibfield  {title}
  {\enquote {\bibinfo {title} {Atrial topology for a unified understanding of
  typical and atypical flutter},}\ }\href {\doibase 10.1161/CIRCEP.124.013102}
  {\bibfield  {journal} {\bibinfo  {journal} {Circ Arrhythm Electrophysiol}\
  }\textbf {\bibinfo {volume} {17}},\ \bibinfo {pages} {e013102} (\bibinfo
  {year} {2024})}\BibitemShut {NoStop}%
\bibitem [{\citenamefont {Van Den~Abeele}\ \emph {et~al.}(2025)\citenamefont
  {Van Den~Abeele}, \citenamefont {Lootens}, \citenamefont {Verstraeten},
  \citenamefont {Bezerra}, \citenamefont {Okenov}, \citenamefont
  {Nezlobinskii}, \citenamefont {Van~Nieuwenhuize}, \citenamefont {Hendrickx},\
  and\ \citenamefont {Vandersickel}}]{Nele2025}%
  \BibitemOpen
  \bibfield  {author} {\bibinfo {author} {\bibfnamefont {Robin}\ \bibnamefont
  {Van Den~Abeele}}, \bibinfo {author} {\bibfnamefont {Sebastiaan}\
  \bibnamefont {Lootens}}, \bibinfo {author} {\bibfnamefont {Bjorn}\
  \bibnamefont {Verstraeten}}, \bibinfo {author} {\bibfnamefont
  {Arthur~Santos}\ \bibnamefont {Bezerra}}, \bibinfo {author} {\bibfnamefont
  {Arstanbek}\ \bibnamefont {Okenov}}, \bibinfo {author} {\bibfnamefont
  {Timur}\ \bibnamefont {Nezlobinskii}}, \bibinfo {author} {\bibfnamefont
  {Viktor}\ \bibnamefont {Van~Nieuwenhuize}}, \bibinfo {author} {\bibfnamefont
  {Sander}\ \bibnamefont {Hendrickx}}, \ and\ \bibinfo {author} {\bibfnamefont
  {Nele}\ \bibnamefont {Vandersickel}},\ }\bibfield  {title} {\enquote
  {\bibinfo {title} {Paired reentries maintain ventricular tachycardia: a
  topological analysis of arrhythmic mechanisms using the index theorem},}\
  }\href {\doibase 10.3389/fnetp.2025.1638085} {\bibfield  {journal} {\bibinfo
  {journal} {Front. Netw. Physiol.}\ }\textbf {\bibinfo {volume} {5}},\
  \bibinfo {pages} {1638085} (\bibinfo {year} {2025})}\BibitemShut {NoStop}%
\bibitem [{\citenamefont {Peyret}\ \emph {et~al.}(2019)\citenamefont {Peyret},
  \citenamefont {Mueller}, \citenamefont {d’Alessandro}, \citenamefont
  {Begnaud}, \citenamefont {Marcq}, \citenamefont {Mège}, \citenamefont
  {Yeomans}, \citenamefont {Doostmohammadi},\ and\ \citenamefont
  {Ladoux}}]{Peyret-oscillations-waves19}%
  \BibitemOpen
  \bibfield  {author} {\bibinfo {author} {\bibfnamefont {Grégoire}\
  \bibnamefont {Peyret}}, \bibinfo {author} {\bibfnamefont {Romain}\
  \bibnamefont {Mueller}}, \bibinfo {author} {\bibfnamefont {Joseph}\
  \bibnamefont {d’Alessandro}}, \bibinfo {author} {\bibfnamefont {Simon}\
  \bibnamefont {Begnaud}}, \bibinfo {author} {\bibfnamefont {Philippe}\
  \bibnamefont {Marcq}}, \bibinfo {author} {\bibfnamefont {René-Marc}\
  \bibnamefont {Mège}}, \bibinfo {author} {\bibfnamefont {Julia~M.}\
  \bibnamefont {Yeomans}}, \bibinfo {author} {\bibfnamefont {Amin}\
  \bibnamefont {Doostmohammadi}}, \ and\ \bibinfo {author} {\bibfnamefont
  {Benoît}\ \bibnamefont {Ladoux}},\ }\bibfield  {title} {\enquote {\bibinfo
  {title} {Sustained oscillations of epithelial cell sheets},}\ }\href
  {\doibase https://doi.org/10.1016/j.bpj.2019.06.013} {\bibfield  {journal}
  {\bibinfo  {journal} {Biophys. J.}\ }\textbf {\bibinfo {volume} {117}},\
  \bibinfo {pages} {464--478} (\bibinfo {year} {2019})}\BibitemShut {NoStop}%
\bibitem [{\citenamefont {Xu}\ \emph {et~al.}(2015)\citenamefont {Xu},
  \citenamefont {Menon}, \citenamefont {Singh}, \citenamefont {Garnier},
  \citenamefont {Sinha},\ and\ \citenamefont {Pumir}}]{Xu2015}%
  \BibitemOpen
  \bibfield  {author} {\bibinfo {author} {\bibfnamefont {Jinshan}\ \bibnamefont
  {Xu}}, \bibinfo {author} {\bibfnamefont {Shakti~N.}\ \bibnamefont {Menon}},
  \bibinfo {author} {\bibfnamefont {Rajeev}\ \bibnamefont {Singh}}, \bibinfo
  {author} {\bibfnamefont {Nicolas~B.}\ \bibnamefont {Garnier}}, \bibinfo
  {author} {\bibfnamefont {Sitabhra}\ \bibnamefont {Sinha}}, \ and\ \bibinfo
  {author} {\bibfnamefont {Alain}\ \bibnamefont {Pumir}},\ }\bibfield  {title}
  {\enquote {\bibinfo {title} {The role of cellular coupling in the spontaneous
  generation of electrical activity in uterine tissue},}\ }\href {\doibase
  10.1371/journal.pone.0118443} {\bibfield  {journal} {\bibinfo  {journal}
  {PLoS ONE}\ }\textbf {\bibinfo {volume} {10}},\ \bibinfo {pages} {1--23}
  (\bibinfo {year} {2015})}\BibitemShut {NoStop}%
\bibitem [{\citenamefont {Myers}\ and\ \citenamefont
  {Elad}(2017{\natexlab{a}})}]{Myers2017}%
  \BibitemOpen
  \bibfield  {author} {\bibinfo {author} {\bibfnamefont {K.~M.}\ \bibnamefont
  {Myers}}\ and\ \bibinfo {author} {\bibfnamefont {D.}~\bibnamefont {Elad}},\
  }\bibfield  {title} {\enquote {\bibinfo {title} {Biomechanics of the human
  uterus},}\ }\href {\doibase 10.1002/wsbm.1388} {\bibfield  {journal}
  {\bibinfo  {journal} {WIREs Syst. Biol. Med.}\ }\textbf {\bibinfo {volume}
  {9}},\ \bibinfo {pages} {e1388} (\bibinfo {year}
  {2017}{\natexlab{a}})}\BibitemShut {NoStop}%
\bibitem [{\citenamefont {Maeda}(2013)}]{Maeda2013UterineCI}%
  \BibitemOpen
  \bibfield  {author} {\bibinfo {author} {\bibfnamefont {Kazuo}\ \bibnamefont
  {Maeda}},\ }\bibfield  {title} {\enquote {\bibinfo {title} {Uterine
  contractions in normal labor developed by a positive feed-backand
  oscillation},}\ }\href {https://api.semanticscholar.org/CorpusID:18075360}
  {\bibfield  {journal} {\bibinfo  {journal} {JHMI}\ }\textbf {\bibinfo
  {volume} {4}},\ \bibinfo {pages} {1--3} (\bibinfo {year} {2013})}\BibitemShut
  {NoStop}%
\bibitem [{\citenamefont {Myers}\ and\ \citenamefont
  {Elad}(2017{\natexlab{b}})}]{Elad2017}%
  \BibitemOpen
  \bibfield  {author} {\bibinfo {author} {\bibfnamefont {Kristin~M.}\
  \bibnamefont {Myers}}\ and\ \bibinfo {author} {\bibfnamefont {David}\
  \bibnamefont {Elad}},\ }\bibfield  {title} {\enquote {\bibinfo {title}
  {Biomechanics of the human uterus},}\ }\href {\doibase
  https://doi.org/10.1002/wsbm.1388} {\bibfield  {journal} {\bibinfo  {journal}
  {WIREs Syst. Biol. Med.}\ }\textbf {\bibinfo {volume} {9}},\ \bibinfo {pages}
  {e1388} (\bibinfo {year} {2017}{\natexlab{b}})}\BibitemShut {NoStop}%
\bibitem [{\citenamefont {Kruse}\ \emph {et~al.}(2004)\citenamefont {Kruse},
  \citenamefont {Joanny}, \citenamefont {J\"ulicher}, \citenamefont {Prost},\
  and\ \citenamefont {Sekimoto}}]{Kruse...Sekimoto2004}%
  \BibitemOpen
  \bibfield  {author} {\bibinfo {author} {\bibfnamefont {K.}~\bibnamefont
  {Kruse}}, \bibinfo {author} {\bibfnamefont {J.~F.}\ \bibnamefont {Joanny}},
  \bibinfo {author} {\bibfnamefont {F.}~\bibnamefont {J\"ulicher}}, \bibinfo
  {author} {\bibfnamefont {J.}~\bibnamefont {Prost}}, \ and\ \bibinfo {author}
  {\bibfnamefont {K.}~\bibnamefont {Sekimoto}},\ }\bibfield  {title} {\enquote
  {\bibinfo {title} {Asters, vortices, and rotating spirals in active gels of
  polar filaments},}\ }\href {\doibase 10.1103/PhysRevLett.92.078101}
  {\bibfield  {journal} {\bibinfo  {journal} {Phys. Rev. Lett.}\ }\textbf
  {\bibinfo {volume} {92}},\ \bibinfo {pages} {078101} (\bibinfo {year}
  {2004})}\BibitemShut {NoStop}%
\bibitem [{\citenamefont {Joanny}\ and\ \citenamefont
  {Prost}(2009)}]{Joanny2009}%
  \BibitemOpen
  \bibfield  {author} {\bibinfo {author} {\bibfnamefont {Jean‐François}\
  \bibnamefont {Joanny}}\ and\ \bibinfo {author} {\bibfnamefont {Jacques}\
  \bibnamefont {Prost}},\ }\bibfield  {title} {\enquote {\bibinfo {title}
  {Active gels as a description of the actin‐myosin cytoskeleton},}\ }\href
  {\doibase 10.2976/1.3054712} {\bibfield  {journal} {\bibinfo  {journal} {HFSP
  J.}\ }\textbf {\bibinfo {volume} {3}},\ \bibinfo {pages} {94--104} (\bibinfo
  {year} {2009})}\BibitemShut {NoStop}%
\bibitem [{\citenamefont {Ranft}\ \emph {et~al.}(2010)\citenamefont {Ranft},
  \citenamefont {Basan}, \citenamefont {Elgeti}, \citenamefont {Joanny},
  \citenamefont {Prost},\ and\ \citenamefont {Jülicher}}]{Jonas2010}%
  \BibitemOpen
  \bibfield  {author} {\bibinfo {author} {\bibfnamefont {Jonas}\ \bibnamefont
  {Ranft}}, \bibinfo {author} {\bibfnamefont {Markus}\ \bibnamefont {Basan}},
  \bibinfo {author} {\bibfnamefont {Jens}\ \bibnamefont {Elgeti}}, \bibinfo
  {author} {\bibfnamefont {Jean-François}\ \bibnamefont {Joanny}}, \bibinfo
  {author} {\bibfnamefont {Jacques}\ \bibnamefont {Prost}}, \ and\ \bibinfo
  {author} {\bibfnamefont {Frank}\ \bibnamefont {Jülicher}},\ }\bibfield
  {title} {\enquote {\bibinfo {title} {Fluidization of tissues by cell division
  and apoptosis},}\ }\href {\doibase 10.1073/pnas.1011086107} {\bibfield
  {journal} {\bibinfo  {journal} {Proc. Natl. Acad. Sci. U. S. A.}\ }\textbf
  {\bibinfo {volume} {107}},\ \bibinfo {pages} {20863--20868} (\bibinfo {year}
  {2010})}\BibitemShut {NoStop}%
\bibitem [{\citenamefont {Prost}\ \emph {et~al.}(2015)\citenamefont {Prost},
  \citenamefont {J{\"u}licher},\ and\ \citenamefont {Joanny}}]{Prost2015}%
  \BibitemOpen
  \bibfield  {author} {\bibinfo {author} {\bibfnamefont {J.}~\bibnamefont
  {Prost}}, \bibinfo {author} {\bibfnamefont {F.}~\bibnamefont {J{\"u}licher}},
  \ and\ \bibinfo {author} {\bibfnamefont {J.-F.}\ \bibnamefont {Joanny}},\
  }\bibfield  {title} {\enquote {\bibinfo {title} {Active gel physics},}\
  }\href {\doibase 10.1038/nphys3224} {\bibfield  {journal} {\bibinfo
  {journal} {Nat. Phys.}\ }\textbf {\bibinfo {volume} {11}},\ \bibinfo {pages}
  {111--117} (\bibinfo {year} {2015})}\BibitemShut {NoStop}%
\bibitem [{\citenamefont {Tlili}\ \emph {et~al.}(2015)\citenamefont {Tlili},
  \citenamefont {Gay}, \citenamefont {Graner}, \citenamefont {Marcq},
  \citenamefont {Molino},\ and\ \citenamefont
  {Saramito}}]{tlili2015colloquium}%
  \BibitemOpen
  \bibfield  {author} {\bibinfo {author} {\bibfnamefont {Sham}\ \bibnamefont
  {Tlili}}, \bibinfo {author} {\bibfnamefont {Cyprien}\ \bibnamefont {Gay}},
  \bibinfo {author} {\bibfnamefont {Fran{\c{c}}ois}\ \bibnamefont {Graner}},
  \bibinfo {author} {\bibfnamefont {Philippe}\ \bibnamefont {Marcq}}, \bibinfo
  {author} {\bibfnamefont {Fran{\c{c}}ois}\ \bibnamefont {Molino}}, \ and\
  \bibinfo {author} {\bibfnamefont {Pierre}\ \bibnamefont {Saramito}},\
  }\bibfield  {title} {\enquote {\bibinfo {title} {Colloquium: Mechanical
  formalisms for tissue dynamics},}\ }\href {\doibase
  10.1140/epje/i2015-15033-4} {\bibfield  {journal} {\bibinfo  {journal} {Eur.
  Phys. J. E}\ }\textbf {\bibinfo {volume} {38}},\ \bibinfo {pages} {33}
  (\bibinfo {year} {2015})}\BibitemShut {NoStop}%
\bibitem [{\citenamefont {Ishihara}\ \emph {et~al.}(2017)\citenamefont
  {Ishihara}, \citenamefont {Marcq},\ and\ \citenamefont
  {Sugimura}}]{ishihara-pre2017}%
  \BibitemOpen
  \bibfield  {author} {\bibinfo {author} {\bibfnamefont {Shuji}\ \bibnamefont
  {Ishihara}}, \bibinfo {author} {\bibfnamefont {Philippe}\ \bibnamefont
  {Marcq}}, \ and\ \bibinfo {author} {\bibfnamefont {Kaoru}\ \bibnamefont
  {Sugimura}},\ }\bibfield  {title} {\enquote {\bibinfo {title} {From cells to
  tissue: A continuum model of epithelial mechanics},}\ }\href {\doibase
  10.1103/PhysRevE.96.022418} {\bibfield  {journal} {\bibinfo  {journal} {Phys.
  Rev. E}\ }\textbf {\bibinfo {volume} {96}},\ \bibinfo {pages} {022418}
  (\bibinfo {year} {2017})}\BibitemShut {NoStop}%
\bibitem [{\citenamefont {Czajkowski}\ \emph {et~al.}(2018)\citenamefont
  {Czajkowski}, \citenamefont {Bi}, \citenamefont {Manning},\ and\
  \citenamefont {Marchetti}}]{C8SM00446C}%
  \BibitemOpen
  \bibfield  {author} {\bibinfo {author} {\bibfnamefont {Michael}\ \bibnamefont
  {Czajkowski}}, \bibinfo {author} {\bibfnamefont {Dapeng}\ \bibnamefont {Bi}},
  \bibinfo {author} {\bibfnamefont {M.~Lisa}\ \bibnamefont {Manning}}, \ and\
  \bibinfo {author} {\bibfnamefont {M.~Cristina}\ \bibnamefont {Marchetti}},\
  }\bibfield  {title} {\enquote {\bibinfo {title} {Hydrodynamics of
  shape-driven rigidity transitions in motile tissues},}\ }\href {\doibase
  10.1039/C8SM00446C} {\bibfield  {journal} {\bibinfo  {journal} {Soft Matter}\
  }\textbf {\bibinfo {volume} {14}},\ \bibinfo {pages} {5628--5642} (\bibinfo
  {year} {2018})}\BibitemShut {NoStop}%
\bibitem [{\citenamefont {Hernandez}\ and\ \citenamefont
  {Marchetti}(2021)}]{hernandez2021}%
  \BibitemOpen
  \bibfield  {author} {\bibinfo {author} {\bibfnamefont {Arthur}\ \bibnamefont
  {Hernandez}}\ and\ \bibinfo {author} {\bibfnamefont {M.~Cristina}\
  \bibnamefont {Marchetti}},\ }\bibfield  {title} {\enquote {\bibinfo {title}
  {Poisson-bracket formulation of the dynamics of fluids of deformable
  particles},}\ }\href {\doibase 10.1103/PhysRevE.103.032612} {\bibfield
  {journal} {\bibinfo  {journal} {Phys. Rev. E}\ }\textbf {\bibinfo {volume}
  {103}},\ \bibinfo {pages} {032612} (\bibinfo {year} {2021})}\BibitemShut
  {NoStop}%
\bibitem [{\citenamefont {Grossman}\ and\ \citenamefont
  {Joanny}(2022)}]{grossman2022instabilities}%
  \BibitemOpen
  \bibfield  {author} {\bibinfo {author} {\bibfnamefont {Doron}\ \bibnamefont
  {Grossman}}\ and\ \bibinfo {author} {\bibfnamefont {Jean-Francois}\
  \bibnamefont {Joanny}},\ }\bibfield  {title} {\enquote {\bibinfo {title}
  {Instabilities and geometry of growing tissues},}\ }\href {\doibase
  10.1103/PhysRevLett.129.048102} {\bibfield  {journal} {\bibinfo  {journal}
  {Phys. Rev. Lett.}\ }\textbf {\bibinfo {volume} {129}},\ \bibinfo {pages}
  {048102} (\bibinfo {year} {2022})}\BibitemShut {NoStop}%
\bibitem [{\citenamefont {Bechinger}\ \emph {et~al.}(2016)\citenamefont
  {Bechinger}, \citenamefont {Di~Leonardo}, \citenamefont {L\"owen},
  \citenamefont {Reichhardt}, \citenamefont {Volpe},\ and\ \citenamefont
  {Volpe}}]{active-matter-review-bechinger}%
  \BibitemOpen
  \bibfield  {author} {\bibinfo {author} {\bibfnamefont {Clemens}\ \bibnamefont
  {Bechinger}}, \bibinfo {author} {\bibfnamefont {Roberto}\ \bibnamefont
  {Di~Leonardo}}, \bibinfo {author} {\bibfnamefont {Hartmut}\ \bibnamefont
  {L\"owen}}, \bibinfo {author} {\bibfnamefont {Charles}\ \bibnamefont
  {Reichhardt}}, \bibinfo {author} {\bibfnamefont {Giorgio}\ \bibnamefont
  {Volpe}}, \ and\ \bibinfo {author} {\bibfnamefont {Giovanni}\ \bibnamefont
  {Volpe}},\ }\bibfield  {title} {\enquote {\bibinfo {title} {Active particles
  in complex and crowded environments},}\ }\href {\doibase
  10.1103/RevModPhys.88.045006} {\bibfield  {journal} {\bibinfo  {journal}
  {Rev. Mod. Phys.}\ }\textbf {\bibinfo {volume} {88}},\ \bibinfo {pages}
  {045006} (\bibinfo {year} {2016})}\BibitemShut {NoStop}%
\bibitem [{\citenamefont {Marchetti}\ \emph {et~al.}(2013)\citenamefont
  {Marchetti}, \citenamefont {Joanny}, \citenamefont {Ramaswamy}, \citenamefont
  {Liverpool}, \citenamefont {Prost}, \citenamefont {Rao},\ and\ \citenamefont
  {Simha}}]{active-matter-marchetti}%
  \BibitemOpen
  \bibfield  {author} {\bibinfo {author} {\bibfnamefont {M.~C.}\ \bibnamefont
  {Marchetti}}, \bibinfo {author} {\bibfnamefont {J.~F.}\ \bibnamefont
  {Joanny}}, \bibinfo {author} {\bibfnamefont {S.}~\bibnamefont {Ramaswamy}},
  \bibinfo {author} {\bibfnamefont {T.~B.}\ \bibnamefont {Liverpool}}, \bibinfo
  {author} {\bibfnamefont {J.}~\bibnamefont {Prost}}, \bibinfo {author}
  {\bibfnamefont {Madan}\ \bibnamefont {Rao}}, \ and\ \bibinfo {author}
  {\bibfnamefont {R.~Aditi}\ \bibnamefont {Simha}},\ }\bibfield  {title}
  {\enquote {\bibinfo {title} {Hydrodynamics of soft active matter},}\ }\href
  {\doibase 10.1103/RevModPhys.85.1143} {\bibfield  {journal} {\bibinfo
  {journal} {Rev. Mod. Phys.}\ }\textbf {\bibinfo {volume} {85}},\ \bibinfo
  {pages} {1143--1189} (\bibinfo {year} {2013})}\BibitemShut {NoStop}%
\bibitem [{\citenamefont {Fodor}\ and\ \citenamefont
  {Marchetti}(2018)}]{Marchetti2018}%
  \BibitemOpen
  \bibfield  {author} {\bibinfo {author} {\bibfnamefont {\'E}\ \bibnamefont
  {Fodor}}\ and\ \bibinfo {author} {\bibfnamefont {M.~Cristina}\ \bibnamefont
  {Marchetti}},\ }\bibfield  {title} {\enquote {\bibinfo {title} {The
  statistical physics of active matter: From self-catalytic colloids to living
  cells},}\ }\href {\doibase https://doi.org/10.1016/j.physa.2017.12.137}
  {\bibfield  {journal} {\bibinfo  {journal} {Physica A}\ }\textbf {\bibinfo
  {volume} {504}},\ \bibinfo {pages} {106--120} (\bibinfo {year}
  {2018})}\BibitemShut {NoStop}%
\bibitem [{\citenamefont {Chaté}(2020)}]{Chate2020}%
  \BibitemOpen
  \bibfield  {author} {\bibinfo {author} {\bibfnamefont {Hugues}\ \bibnamefont
  {Chaté}},\ }\bibfield  {title} {\enquote {\bibinfo {title} {Dry aligning
  dilute active matter},}\ }\href {\doibase
  https://doi.org/10.1146/annurev-conmatphys-031119-050752} {\bibfield
  {journal} {\bibinfo  {journal} {Annu. Rev. Condens. Matter Phys.}\ }\textbf
  {\bibinfo {volume} {11}},\ \bibinfo {pages} {189--212} (\bibinfo {year}
  {2020})}\BibitemShut {NoStop}%
\bibitem [{\citenamefont {Cates}\ and\ \citenamefont {Tailleur}(2015)}]{MIPS}%
  \BibitemOpen
  \bibfield  {author} {\bibinfo {author} {\bibfnamefont {Michael~E.}\
  \bibnamefont {Cates}}\ and\ \bibinfo {author} {\bibfnamefont {Julien}\
  \bibnamefont {Tailleur}},\ }\bibfield  {title} {\enquote {\bibinfo {title}
  {Motility-induced phase separation},}\ }\href {\doibase
  10.1146/annurev-conmatphys-031214-014710} {\bibfield  {journal} {\bibinfo
  {journal} {Annu. Rev. Condens. Matter Phys.}\ }\textbf {\bibinfo {volume}
  {6}},\ \bibinfo {pages} {219--244} (\bibinfo {year} {2015})}\BibitemShut
  {NoStop}%
\bibitem [{\citenamefont {Martin}\ \emph {et~al.}(2009)\citenamefont {Martin},
  \citenamefont {Kaschube},\ and\ \citenamefont
  {Wieschaus}}]{martin2009pulsed}%
  \BibitemOpen
  \bibfield  {author} {\bibinfo {author} {\bibfnamefont {Adam~C.}\ \bibnamefont
  {Martin}}, \bibinfo {author} {\bibfnamefont {Matthias}\ \bibnamefont
  {Kaschube}}, \ and\ \bibinfo {author} {\bibfnamefont {Eric~F.}\ \bibnamefont
  {Wieschaus}},\ }\bibfield  {title} {\enquote {\bibinfo {title} {Pulsed
  contractions of an actin--myosin network drive apical constriction},}\ }\href
  {\doibase 10.1038/nature07522} {\bibfield  {journal} {\bibinfo  {journal}
  {Nature}\ }\textbf {\bibinfo {volume} {457}},\ \bibinfo {pages} {495--499}
  (\bibinfo {year} {2009})}\BibitemShut {NoStop}%
\bibitem [{\citenamefont {Deforet}\ \emph {et~al.}(2014)\citenamefont
  {Deforet}, \citenamefont {Hakim}, \citenamefont {Yevick}, \citenamefont
  {Duclos},\ and\ \citenamefont {Silberzan}}]{Hakim-oscillations-NatComm14}%
  \BibitemOpen
  \bibfield  {author} {\bibinfo {author} {\bibfnamefont {M.}~\bibnamefont
  {Deforet}}, \bibinfo {author} {\bibfnamefont {V.}~\bibnamefont {Hakim}},
  \bibinfo {author} {\bibfnamefont {H.~G.}\ \bibnamefont {Yevick}}, \bibinfo
  {author} {\bibfnamefont {G.}~\bibnamefont {Duclos}}, \ and\ \bibinfo {author}
  {\bibfnamefont {P.}~\bibnamefont {Silberzan}},\ }\bibfield  {title} {\enquote
  {\bibinfo {title} {Emergence of collective modes and tri-dimensional
  structures from epithelial confinement},}\ }\href {\doibase
  10.1038/ncomms4747} {\bibfield  {journal} {\bibinfo  {journal} {Nat.
  Commun.}\ }\textbf {\bibinfo {volume} {5}},\ \bibinfo {pages} {3747}
  (\bibinfo {year} {2014})}\BibitemShut {NoStop}%
\bibitem [{\citenamefont {Serra-Picamal}\ \emph {et~al.}(2012)\citenamefont
  {Serra-Picamal}, \citenamefont {Conte}, \citenamefont {Vincent},
  \citenamefont {Anon}, \citenamefont {Tambe}, \citenamefont {Bazellieres},
  \citenamefont {Butler}, \citenamefont {Fredberg},\ and\ \citenamefont
  {Trepat}}]{Serra-Picamal-Natphys2012}%
  \BibitemOpen
  \bibfield  {author} {\bibinfo {author} {\bibfnamefont {Xavier}\ \bibnamefont
  {Serra-Picamal}}, \bibinfo {author} {\bibfnamefont {Vito}\ \bibnamefont
  {Conte}}, \bibinfo {author} {\bibfnamefont {Romaric}\ \bibnamefont
  {Vincent}}, \bibinfo {author} {\bibfnamefont {Ester}\ \bibnamefont {Anon}},
  \bibinfo {author} {\bibfnamefont {Dhananjay~T.}\ \bibnamefont {Tambe}},
  \bibinfo {author} {\bibfnamefont {Elsa}\ \bibnamefont {Bazellieres}},
  \bibinfo {author} {\bibfnamefont {James~P.}\ \bibnamefont {Butler}}, \bibinfo
  {author} {\bibfnamefont {Jeffrey~J.}\ \bibnamefont {Fredberg}}, \ and\
  \bibinfo {author} {\bibfnamefont {Xavier}\ \bibnamefont {Trepat}},\
  }\bibfield  {title} {\enquote {\bibinfo {title} {Mechanical waves during
  tissue expansion},}\ }\href {\doibase 10.1038/nphys2355} {\bibfield
  {journal} {\bibinfo  {journal} {Nat. Phys.}\ }\textbf {\bibinfo {volume}
  {8}},\ \bibinfo {pages} {628--634} (\bibinfo {year} {2012})}\BibitemShut
  {NoStop}%
\bibitem [{\citenamefont {Zaritsky}\ \emph {et~al.}(2014)\citenamefont
  {Zaritsky}, \citenamefont {Kaplan}, \citenamefont {Hecht}, \citenamefont
  {Natan}, \citenamefont {Wolf}, \citenamefont {Gov}, \citenamefont
  {Ben-Jacob},\ and\ \citenamefont {Tsarfaty}}]{Zaritsky-PLoS}%
  \BibitemOpen
  \bibfield  {author} {\bibinfo {author} {\bibfnamefont {Assaf}\ \bibnamefont
  {Zaritsky}}, \bibinfo {author} {\bibfnamefont {Doron}\ \bibnamefont
  {Kaplan}}, \bibinfo {author} {\bibfnamefont {Inbal}\ \bibnamefont {Hecht}},
  \bibinfo {author} {\bibfnamefont {Sari}\ \bibnamefont {Natan}}, \bibinfo
  {author} {\bibfnamefont {Lior}\ \bibnamefont {Wolf}}, \bibinfo {author}
  {\bibfnamefont {Nir~S.}\ \bibnamefont {Gov}}, \bibinfo {author}
  {\bibfnamefont {Eshel}\ \bibnamefont {Ben-Jacob}}, \ and\ \bibinfo {author}
  {\bibfnamefont {Ilan}\ \bibnamefont {Tsarfaty}},\ }\bibfield  {title}
  {\enquote {\bibinfo {title} {Propagating waves of directionality and
  coordination orchestrate collective cell migration},}\ }\href {\doibase
  10.1371/journal.pcbi.1003747} {\bibfield  {journal} {\bibinfo  {journal}
  {PLoS Comput. Biol.}\ }\textbf {\bibinfo {volume} {10}},\ \bibinfo {pages}
  {e1003747} (\bibinfo {year} {2014})}\BibitemShut {NoStop}%
\bibitem [{\citenamefont {Banerjee}\ \emph {et~al.}(2015)\citenamefont
  {Banerjee}, \citenamefont {Utuje},\ and\ \citenamefont
  {Marchetti}}]{Banerjee-Marchetti-PRL2015}%
  \BibitemOpen
  \bibfield  {author} {\bibinfo {author} {\bibfnamefont {Shiladitya}\
  \bibnamefont {Banerjee}}, \bibinfo {author} {\bibfnamefont {Kazage J.~C.}\
  \bibnamefont {Utuje}}, \ and\ \bibinfo {author} {\bibfnamefont {M.~Cristina}\
  \bibnamefont {Marchetti}},\ }\bibfield  {title} {\enquote {\bibinfo {title}
  {Propagating stress waves during epithelial expansion},}\ }\href {\doibase
  10.1103/PhysRevLett.114.228101} {\bibfield  {journal} {\bibinfo  {journal}
  {Phys. Rev. Lett.}\ }\textbf {\bibinfo {volume} {114}},\ \bibinfo {pages}
  {228101} (\bibinfo {year} {2015})}\BibitemShut {NoStop}%
\bibitem [{\citenamefont {Tlili}\ \emph {et~al.}(2018)\citenamefont {Tlili},
  \citenamefont {Gauquelin}, \citenamefont {Li}, \citenamefont {Cardoso},
  \citenamefont {Ladoux}, \citenamefont {Delanoë-Ayari},\ and\ \citenamefont
  {Graner}}]{tlili-RSoc18}%
  \BibitemOpen
  \bibfield  {author} {\bibinfo {author} {\bibfnamefont {Sham}\ \bibnamefont
  {Tlili}}, \bibinfo {author} {\bibfnamefont {Estelle}\ \bibnamefont
  {Gauquelin}}, \bibinfo {author} {\bibfnamefont {Brigitte}\ \bibnamefont
  {Li}}, \bibinfo {author} {\bibfnamefont {Olivier}\ \bibnamefont {Cardoso}},
  \bibinfo {author} {\bibfnamefont {Benoît}\ \bibnamefont {Ladoux}}, \bibinfo
  {author} {\bibfnamefont {Hélène}\ \bibnamefont {Delanoë-Ayari}}, \ and\
  \bibinfo {author} {\bibfnamefont {Fran\c{c}ois}\ \bibnamefont {Graner}},\
  }\bibfield  {title} {\enquote {\bibinfo {title} {Collective cell migration
  without proliferation: density determines cell velocity and wave velocity},}\
  }\href {\doibase 10.1098/rsos.172421} {\bibfield  {journal} {\bibinfo
  {journal} {R. Soc. Open Sci.}\ }\textbf {\bibinfo {volume} {5}},\ \bibinfo
  {pages} {172421} (\bibinfo {year} {2018})}\BibitemShut {NoStop}%
\bibitem [{\citenamefont {Petrolli}\ \emph {et~al.}(2019)\citenamefont
  {Petrolli}, \citenamefont {Le~Goff}, \citenamefont {Tadrous}, \citenamefont
  {Martens}, \citenamefont {Allier}, \citenamefont {Mandula}, \citenamefont
  {Herv\'e}, \citenamefont {Henkes}, \citenamefont {Sknepnek}, \citenamefont
  {Boudou}, \citenamefont {Cappello},\ and\ \citenamefont
  {Balland}}]{Petrolli-PRL2019}%
  \BibitemOpen
  \bibfield  {author} {\bibinfo {author} {\bibfnamefont {Vanni}\ \bibnamefont
  {Petrolli}}, \bibinfo {author} {\bibfnamefont {Magali}\ \bibnamefont
  {Le~Goff}}, \bibinfo {author} {\bibfnamefont {Monika}\ \bibnamefont
  {Tadrous}}, \bibinfo {author} {\bibfnamefont {Kirsten}\ \bibnamefont
  {Martens}}, \bibinfo {author} {\bibfnamefont {C\'edric}\ \bibnamefont
  {Allier}}, \bibinfo {author} {\bibfnamefont {Ondrej}\ \bibnamefont
  {Mandula}}, \bibinfo {author} {\bibfnamefont {Lionel}\ \bibnamefont
  {Herv\'e}}, \bibinfo {author} {\bibfnamefont {Silke}\ \bibnamefont {Henkes}},
  \bibinfo {author} {\bibfnamefont {Rastko}\ \bibnamefont {Sknepnek}}, \bibinfo
  {author} {\bibfnamefont {Thomas}\ \bibnamefont {Boudou}}, \bibinfo {author}
  {\bibfnamefont {Giovanni}\ \bibnamefont {Cappello}}, \ and\ \bibinfo {author}
  {\bibfnamefont {Martial}\ \bibnamefont {Balland}},\ }\bibfield  {title}
  {\enquote {\bibinfo {title} {Confinement-induced transition between wavelike
  collective cell migration modes},}\ }\href {\doibase
  10.1103/PhysRevLett.122.168101} {\bibfield  {journal} {\bibinfo  {journal}
  {Phys. Rev. Lett.}\ }\textbf {\bibinfo {volume} {122}},\ \bibinfo {pages}
  {168101} (\bibinfo {year} {2019})}\BibitemShut {NoStop}%
\bibitem [{\citenamefont {Hino}\ \emph {et~al.}(2020)\citenamefont {Hino},
  \citenamefont {Rossetti}, \citenamefont {Marín-Llauradó}, \citenamefont
  {Aoki}, \citenamefont {Trepat}, \citenamefont {Matsuda},\ and\ \citenamefont
  {Hirashima}}]{Hino-DevCell2020}%
  \BibitemOpen
  \bibfield  {author} {\bibinfo {author} {\bibfnamefont {Naoya}\ \bibnamefont
  {Hino}}, \bibinfo {author} {\bibfnamefont {Leone}\ \bibnamefont {Rossetti}},
  \bibinfo {author} {\bibfnamefont {Ariadna}\ \bibnamefont {Marín-Llauradó}},
  \bibinfo {author} {\bibfnamefont {Kazuhiro}\ \bibnamefont {Aoki}}, \bibinfo
  {author} {\bibfnamefont {Xavier}\ \bibnamefont {Trepat}}, \bibinfo {author}
  {\bibfnamefont {Michiyuki}\ \bibnamefont {Matsuda}}, \ and\ \bibinfo {author}
  {\bibfnamefont {Tsuyoshi}\ \bibnamefont {Hirashima}},\ }\bibfield  {title}
  {\enquote {\bibinfo {title} {Erk-mediated mechanochemical waves direct
  collective cell polarization},}\ }\href {\doibase
  10.1016/j.devcel.2020.05.011} {\bibfield  {journal} {\bibinfo  {journal}
  {Dev. Cell}\ }\textbf {\bibinfo {volume} {53}},\ \bibinfo {pages}
  {646--660.e8} (\bibinfo {year} {2020})}\BibitemShut {NoStop}%
\bibitem [{\citenamefont {Armon}\ \emph {et~al.}(2018)\citenamefont {Armon},
  \citenamefont {Bull}, \citenamefont {Aranda-Diaz},\ and\ \citenamefont
  {Prakash}}]{Armon-PNAS}%
  \BibitemOpen
  \bibfield  {author} {\bibinfo {author} {\bibfnamefont {Shahaf}\ \bibnamefont
  {Armon}}, \bibinfo {author} {\bibfnamefont {Matthew~Storm}\ \bibnamefont
  {Bull}}, \bibinfo {author} {\bibfnamefont {Andres}\ \bibnamefont
  {Aranda-Diaz}}, \ and\ \bibinfo {author} {\bibfnamefont {Manu}\ \bibnamefont
  {Prakash}},\ }\bibfield  {title} {\enquote {\bibinfo {title} {Ultrafast
  epithelial contractions provide insights into contraction speed limits and
  tissue integrity},}\ }\href {\doibase 10.1073/pnas.1802934115} {\bibfield
  {journal} {\bibinfo  {journal} {Proc. Natl. Acad. Sci. U. S. A.}\ }\textbf
  {\bibinfo {volume} {115}} (\bibinfo {year} {2018}),\
  10.1073/pnas.1802934115}\BibitemShut {NoStop}%
\bibitem [{\citenamefont {Young}(1997)}]{Young1997}%
  \BibitemOpen
  \bibfield  {author} {\bibinfo {author} {\bibfnamefont {Roger~C.}\
  \bibnamefont {Young}},\ }\bibfield  {title} {\enquote {\bibinfo {title} {A
  computer model of uterine contractions based on action potential propagation
  and intercellular calcium waves},}\ }\href {\doibase
  https://doi.org/10.1016/S0029-7844(96)00502-9} {\bibfield  {journal}
  {\bibinfo  {journal} {Obstet. Gynecol.}\ }\textbf {\bibinfo {volume} {89}},\
  \bibinfo {pages} {604--608} (\bibinfo {year} {1997})}\BibitemShut {NoStop}%
\bibitem [{\citenamefont {Kruse}\ and\ \citenamefont
  {Riveline}(2011)}]{kruse2011spontaneous}%
  \BibitemOpen
  \bibfield  {author} {\bibinfo {author} {\bibfnamefont {Karsten}\ \bibnamefont
  {Kruse}}\ and\ \bibinfo {author} {\bibfnamefont {Daniel}\ \bibnamefont
  {Riveline}},\ }\bibfield  {title} {\enquote {\bibinfo {title} {Spontaneous
  mechanical oscillations: implications for developing organisms},}\ }\href
  {\doibase https://doi.org/10.1016/B978-0-12-385065-2.00003-7} {\bibfield
  {journal} {\bibinfo  {journal} {Curr. Top. Dev. Biol.}\ }\textbf {\bibinfo
  {volume} {95}},\ \bibinfo {pages} {67--91} (\bibinfo {year}
  {2011})}\BibitemShut {NoStop}%
\bibitem [{\citenamefont {Collinet}\ and\ \citenamefont
  {Lecuit}(2021)}]{collinet2021programmed}%
  \BibitemOpen
  \bibfield  {author} {\bibinfo {author} {\bibfnamefont {Claudio}\ \bibnamefont
  {Collinet}}\ and\ \bibinfo {author} {\bibfnamefont {Thomas}\ \bibnamefont
  {Lecuit}},\ }\bibfield  {title} {\enquote {\bibinfo {title} {Programmed and
  self-organized flow of information during morphogenesis},}\ }\href {\doibase
  10.1038/s41580-020-00318-6} {\bibfield  {journal} {\bibinfo  {journal} {Nat
  Rev Mol Cell Biol.}\ }\textbf {\bibinfo {volume} {22}},\ \bibinfo {pages}
  {245--265} (\bibinfo {year} {2021})}\BibitemShut {NoStop}%
\bibitem [{\citenamefont {Bi}\ \emph {et~al.}(2016)\citenamefont {Bi},
  \citenamefont {Yang}, \citenamefont {Marchetti},\ and\ \citenamefont
  {Manning}}]{PhysRevX.6.021011}%
  \BibitemOpen
  \bibfield  {author} {\bibinfo {author} {\bibfnamefont {Dapeng}\ \bibnamefont
  {Bi}}, \bibinfo {author} {\bibfnamefont {Xingbo}\ \bibnamefont {Yang}},
  \bibinfo {author} {\bibfnamefont {M.~Cristina}\ \bibnamefont {Marchetti}}, \
  and\ \bibinfo {author} {\bibfnamefont {M.~Lisa}\ \bibnamefont {Manning}},\
  }\bibfield  {title} {\enquote {\bibinfo {title} {Motility-driven glass and
  jamming transitions in biological tissues},}\ }\href {\doibase
  10.1103/PhysRevX.6.021011} {\bibfield  {journal} {\bibinfo  {journal} {Phys.
  Rev. X}\ }\textbf {\bibinfo {volume} {6}},\ \bibinfo {pages} {021011}
  (\bibinfo {year} {2016})}\BibitemShut {NoStop}%
\bibitem [{\citenamefont {Manning}(2023)}]{Manning-essay}%
  \BibitemOpen
  \bibfield  {author} {\bibinfo {author} {\bibfnamefont {M.~Lisa}\ \bibnamefont
  {Manning}},\ }\bibfield  {title} {\enquote {\bibinfo {title} {Essay:
  Collections of deformable particles present exciting challenges for soft
  matter and biological physics},}\ }\href {\doibase
  10.1103/PhysRevLett.130.130002} {\bibfield  {journal} {\bibinfo  {journal}
  {Phys. Rev. Lett.}\ }\textbf {\bibinfo {volume} {130}},\ \bibinfo {pages}
  {130002} (\bibinfo {year} {2023})}\BibitemShut {NoStop}%
\bibitem [{\citenamefont {Hopkins}\ \emph {et~al.}(2022)\citenamefont
  {Hopkins}, \citenamefont {Chiang}, \citenamefont {Loewe}, \citenamefont
  {Marenduzzo},\ and\ \citenamefont {Marchetti}}]{PhysRevLett.129.148101}%
  \BibitemOpen
  \bibfield  {author} {\bibinfo {author} {\bibfnamefont {Austin}\ \bibnamefont
  {Hopkins}}, \bibinfo {author} {\bibfnamefont {Michael}\ \bibnamefont
  {Chiang}}, \bibinfo {author} {\bibfnamefont {Benjamin}\ \bibnamefont
  {Loewe}}, \bibinfo {author} {\bibfnamefont {Davide}\ \bibnamefont
  {Marenduzzo}}, \ and\ \bibinfo {author} {\bibfnamefont {M.~Cristina}\
  \bibnamefont {Marchetti}},\ }\bibfield  {title} {\enquote {\bibinfo {title}
  {Local yield and compliance in active cell monolayers},}\ }\href {\doibase
  10.1103/PhysRevLett.129.148101} {\bibfield  {journal} {\bibinfo  {journal}
  {Phys. Rev. Lett.}\ }\textbf {\bibinfo {volume} {129}},\ \bibinfo {pages}
  {148101} (\bibinfo {year} {2022})}\BibitemShut {NoStop}%
\bibitem [{\citenamefont {Zhang}\ and\ \citenamefont
  {Yeomans}(2023)}]{PhysRevLett.130.038202}%
  \BibitemOpen
  \bibfield  {author} {\bibinfo {author} {\bibfnamefont {Guanming}\
  \bibnamefont {Zhang}}\ and\ \bibinfo {author} {\bibfnamefont {Julia~M.}\
  \bibnamefont {Yeomans}},\ }\bibfield  {title} {\enquote {\bibinfo {title}
  {Active forces in confluent cell monolayers},}\ }\href {\doibase
  10.1103/PhysRevLett.130.038202} {\bibfield  {journal} {\bibinfo  {journal}
  {Phys. Rev. Lett.}\ }\textbf {\bibinfo {volume} {130}},\ \bibinfo {pages}
  {038202} (\bibinfo {year} {2023})}\BibitemShut {NoStop}%
\bibitem [{\citenamefont {G{\"o}th}\ and\ \citenamefont
  {Dzubiella}(2025)}]{goth2025}%
  \BibitemOpen
  \bibfield  {author} {\bibinfo {author} {\bibfnamefont {Nils}\ \bibnamefont
  {G{\"o}th}}\ and\ \bibinfo {author} {\bibfnamefont {Joachim}\ \bibnamefont
  {Dzubiella}},\ }\bibfield  {title} {\enquote {\bibinfo {title} {Collective
  chemo-mechanical oscillations and cluster waves in communicating colloids},}\
  }\href {\doibase 0.1038/s42005-025-01983-9} {\bibfield  {journal} {\bibinfo
  {journal} {Commun. Phys.}\ }\textbf {\bibinfo {volume} {8}},\ \bibinfo
  {pages} {65} (\bibinfo {year} {2025})}\BibitemShut {NoStop}%
\bibitem [{\citenamefont {Zhang}\ \emph {et~al.}(2025)\citenamefont {Zhang},
  \citenamefont {Manacorda},\ and\ \citenamefont {Fodor}}]{Zhang_2025}%
  \BibitemOpen
  \bibfield  {author} {\bibinfo {author} {\bibfnamefont {Yiwei}\ \bibnamefont
  {Zhang}}, \bibinfo {author} {\bibfnamefont {Alessandro}\ \bibnamefont
  {Manacorda}}, \ and\ \bibinfo {author} {\bibfnamefont {Étienne}\
  \bibnamefont {Fodor}},\ }\bibfield  {title} {\enquote {\bibinfo {title}
  {Species interconversion of deformable particles yields transient phase
  separation},}\ }\href {\doibase 10.1088/1367-2630/adccf1} {\bibfield
  {journal} {\bibinfo  {journal} {New J. Phys.}\ }\textbf {\bibinfo {volume}
  {27}},\ \bibinfo {pages} {043023} (\bibinfo {year} {2025})}\BibitemShut
  {NoStop}%
\bibitem [{\citenamefont {Farhadifar}\ \emph {et~al.}(2007)\citenamefont
  {Farhadifar}, \citenamefont {Röper}, \citenamefont {Aigouy}, \citenamefont
  {Eaton},\ and\ \citenamefont {Jülicher}}]{farhadifar2007influence}%
  \BibitemOpen
  \bibfield  {author} {\bibinfo {author} {\bibfnamefont {Reza}\ \bibnamefont
  {Farhadifar}}, \bibinfo {author} {\bibfnamefont {Jens-Christian}\
  \bibnamefont {Röper}}, \bibinfo {author} {\bibfnamefont {Benoit}\
  \bibnamefont {Aigouy}}, \bibinfo {author} {\bibfnamefont {Suzanne}\
  \bibnamefont {Eaton}}, \ and\ \bibinfo {author} {\bibfnamefont {Frank}\
  \bibnamefont {Jülicher}},\ }\bibfield  {title} {\enquote {\bibinfo {title}
  {The influence of cell mechanics, cell-cell interactions, and proliferation
  on epithelial packing},}\ }\href {\doibase
  https://doi.org/10.1016/j.cub.2007.11.049} {\bibfield  {journal} {\bibinfo
  {journal} {Curr. Biol.}\ }\textbf {\bibinfo {volume} {17}},\ \bibinfo {pages}
  {2095--2104} (\bibinfo {year} {2007})}\BibitemShut {NoStop}%
\bibitem [{\citenamefont {Staple}\ \emph {et~al.}(2010)\citenamefont {Staple},
  \citenamefont {Farhadifar}, \citenamefont {R{\"o}per}, \citenamefont
  {Aigouy}, \citenamefont {Eaton},\ and\ \citenamefont
  {J{\"u}licher}}]{staple2010mechanics}%
  \BibitemOpen
  \bibfield  {author} {\bibinfo {author} {\bibfnamefont {D.~B.}\ \bibnamefont
  {Staple}}, \bibinfo {author} {\bibfnamefont {R.}~\bibnamefont {Farhadifar}},
  \bibinfo {author} {\bibfnamefont {J.-C.}\ \bibnamefont {R{\"o}per}}, \bibinfo
  {author} {\bibfnamefont {B.}~\bibnamefont {Aigouy}}, \bibinfo {author}
  {\bibfnamefont {S.}~\bibnamefont {Eaton}}, \ and\ \bibinfo {author}
  {\bibfnamefont {F.}~\bibnamefont {J{\"u}licher}},\ }\bibfield  {title}
  {\enquote {\bibinfo {title} {Mechanics and remodelling of cell packings in
  epithelia},}\ }\href {\doibase 10.1140/epje/i2010-10677-0} {\bibfield
  {journal} {\bibinfo  {journal} {Eur. Phys. J. E}\ }\textbf {\bibinfo {volume}
  {33}},\ \bibinfo {pages} {117--127} (\bibinfo {year} {2010})}\BibitemShut
  {NoStop}%
\bibitem [{\citenamefont {Armon}\ \emph {et~al.}(2021)\citenamefont {Armon},
  \citenamefont {Bull}, \citenamefont {Moriel}, \citenamefont {Aharoni},\ and\
  \citenamefont {Prakash}}]{Armon-Comm-phys}%
  \BibitemOpen
  \bibfield  {author} {\bibinfo {author} {\bibfnamefont {Shahaf}\ \bibnamefont
  {Armon}}, \bibinfo {author} {\bibfnamefont {Matthew~S.}\ \bibnamefont
  {Bull}}, \bibinfo {author} {\bibfnamefont {Avraham}\ \bibnamefont {Moriel}},
  \bibinfo {author} {\bibfnamefont {Hillel}\ \bibnamefont {Aharoni}}, \ and\
  \bibinfo {author} {\bibfnamefont {Manu}\ \bibnamefont {Prakash}},\ }\bibfield
   {title} {\enquote {\bibinfo {title} {Modeling epithelial tissues as
  active-elastic sheets reproduce contraction pulses and predict rip
  resistance},}\ }\href {\doibase 10.1038/s42005-021-00712-2} {\bibfield
  {journal} {\bibinfo  {journal} {Commun. Phys.}\ }\textbf {\bibinfo {volume}
  {4}},\ \bibinfo {pages} {216} (\bibinfo {year} {2021})}\BibitemShut {NoStop}%
\bibitem [{\citenamefont {Pérez-Verdugo}\ \emph {et~al.}(2022)\citenamefont
  {Pérez-Verdugo}, \citenamefont {Reig}, \citenamefont {Cerda}, \citenamefont
  {Concha},\ and\ \citenamefont {Soto}}]{Soto2022}%
  \BibitemOpen
  \bibfield  {author} {\bibinfo {author} {\bibfnamefont {Fernanda}\
  \bibnamefont {Pérez-Verdugo}}, \bibinfo {author} {\bibfnamefont {Germán}\
  \bibnamefont {Reig}}, \bibinfo {author} {\bibfnamefont {Mauricio}\
  \bibnamefont {Cerda}}, \bibinfo {author} {\bibfnamefont {Miguel~L.}\
  \bibnamefont {Concha}}, \ and\ \bibinfo {author} {\bibfnamefont {Rodrigo}\
  \bibnamefont {Soto}},\ }\bibfield  {title} {\enquote {\bibinfo {title}
  {Geometrical characterization of active contraction pulses in epithelial
  cells using the two-dimensional vertex model},}\ }\href {\doibase
  10.1098/rsif.2021.0851} {\bibfield  {journal} {\bibinfo  {journal} {J. R.
  Soc. Interface}\ }\textbf {\bibinfo {volume} {19}},\ \bibinfo {pages}
  {20210851} (\bibinfo {year} {2022})}\BibitemShut {NoStop}%
\bibitem [{\citenamefont {Boocock}\ \emph {et~al.}(2023)\citenamefont
  {Boocock}, \citenamefont {Hirashima},\ and\ \citenamefont
  {Hannezo}}]{Hannezo-PRXLife}%
  \BibitemOpen
  \bibfield  {author} {\bibinfo {author} {\bibfnamefont {Daniel}\ \bibnamefont
  {Boocock}}, \bibinfo {author} {\bibfnamefont {Tsuyoshi}\ \bibnamefont
  {Hirashima}}, \ and\ \bibinfo {author} {\bibfnamefont {Edouard}\ \bibnamefont
  {Hannezo}},\ }\bibfield  {title} {\enquote {\bibinfo {title} {Interplay
  between mechanochemical patterning and glassy dynamics in cellular
  monolayers},}\ }\href {\doibase 10.1103/PRXLife.1.013001} {\bibfield
  {journal} {\bibinfo  {journal} {PRX Life}\ }\textbf {\bibinfo {volume} {1}},\
  \bibinfo {pages} {013001} (\bibinfo {year} {2023})}\BibitemShut {NoStop}%
\bibitem [{\citenamefont {P{\'e}rez-Verdugo}\ \emph {et~al.}(2024)\citenamefont
  {P{\'e}rez-Verdugo}, \citenamefont {Banks},\ and\ \citenamefont
  {Banerjee}}]{Shiladitya-arxiv23}%
  \BibitemOpen
  \bibfield  {author} {\bibinfo {author} {\bibfnamefont {Fernanda}\
  \bibnamefont {P{\'e}rez-Verdugo}}, \bibinfo {author} {\bibfnamefont {Samuel}\
  \bibnamefont {Banks}}, \ and\ \bibinfo {author} {\bibfnamefont {Shiladitya}\
  \bibnamefont {Banerjee}},\ }\bibfield  {title} {\enquote {\bibinfo {title}
  {Excitable dynamics driven by mechanical feedback in biological tissues},}\
  }\href {\doibase 10.1038/s42005-024-01661-2} {\bibfield  {journal} {\bibinfo
  {journal} {Commun. Phys.}\ }\textbf {\bibinfo {volume} {7}},\ \bibinfo
  {pages} {167} (\bibinfo {year} {2024})}\BibitemShut {NoStop}%
\bibitem [{\citenamefont {Togashi}(2019)}]{Togashi2019}%
  \BibitemOpen
  \bibfield  {author} {\bibinfo {author} {\bibfnamefont {Yuichi}\ \bibnamefont
  {Togashi}},\ }\bibfield  {title} {\enquote {\bibinfo {title} {Modeling of
  nanomachine/micromachine crowds: Interplay between the internal state and
  surroundings},}\ }\href {\doibase 10.1021/acs.jpcb.8b10633} {\bibfield
  {journal} {\bibinfo  {journal} {J. Phys. Chem. B}\ }\textbf {\bibinfo
  {volume} {123}},\ \bibinfo {pages} {1481--1490} (\bibinfo {year}
  {2019})}\BibitemShut {NoStop}%
\bibitem [{\citenamefont {Zhang}\ and\ \citenamefont
  {Fodor}(2023)}]{Yiwei-Etienne-PAM}%
  \BibitemOpen
  \bibfield  {author} {\bibinfo {author} {\bibfnamefont {Yiwei}\ \bibnamefont
  {Zhang}}\ and\ \bibinfo {author} {\bibfnamefont {\'Etienne}\ \bibnamefont
  {Fodor}},\ }\bibfield  {title} {\enquote {\bibinfo {title} {Pulsating active
  matter},}\ }\href {\doibase 10.1103/PhysRevLett.131.238302} {\bibfield
  {journal} {\bibinfo  {journal} {Phys. Rev. Lett.}\ }\textbf {\bibinfo
  {volume} {131}},\ \bibinfo {pages} {238302} (\bibinfo {year}
  {2023})}\BibitemShut {NoStop}%
\bibitem [{\citenamefont {hua Liu}\ \emph {et~al.}(2024)\citenamefont {hua
  Liu}, \citenamefont {jing Zhu},\ and\ \citenamefont {quan Ai}}]{Liu_2024}%
  \BibitemOpen
  \bibfield  {author} {\bibinfo {author} {\bibfnamefont {Wan}\ \bibnamefont
  {hua Liu}}, \bibinfo {author} {\bibfnamefont {Wei}\ \bibnamefont {jing Zhu}},
  \ and\ \bibinfo {author} {\bibfnamefont {Bao}\ \bibnamefont {quan Ai}},\
  }\bibfield  {title} {\enquote {\bibinfo {title} {Collective motion of
  pulsating active particles in confined structures},}\ }\href {\doibase
  10.1088/1367-2630/ad23a5} {\bibfield  {journal} {\bibinfo  {journal} {New J.
  Phys.}\ }\textbf {\bibinfo {volume} {26}},\ \bibinfo {pages} {023017}
  (\bibinfo {year} {2024})}\BibitemShut {NoStop}%
\bibitem [{\citenamefont {Li}\ \emph {et~al.}(2025)\citenamefont {Li},
  \citenamefont {Lei},\ and\ \citenamefont {Ma}}]{li2024fluidization}%
  \BibitemOpen
  \bibfield  {author} {\bibinfo {author} {\bibfnamefont {Zhu-Qin}\ \bibnamefont
  {Li}}, \bibinfo {author} {\bibfnamefont {Qun-Li}\ \bibnamefont {Lei}}, \ and\
  \bibinfo {author} {\bibfnamefont {Yu-Qiang}\ \bibnamefont {Ma}},\ }\bibfield
  {title} {\enquote {\bibinfo {title} {Fluidization and anomalous density
  fluctuations in 2d voronoi cell tissues with pulsating activity},}\ }\href
  {\doibase 10.1073/pnas.2421518122} {\bibfield  {journal} {\bibinfo  {journal}
  {Proc. Natl. Acad. Sci. U.S.A.}\ }\textbf {\bibinfo {volume} {122}},\
  \bibinfo {pages} {e2421518122} (\bibinfo {year} {2025})}\BibitemShut
  {NoStop}%
\bibitem [{\citenamefont {Pi\~neros}\ and\ \citenamefont
  {Fodor}(2025)}]{pineros2024biased}%
  \BibitemOpen
  \bibfield  {author} {\bibinfo {author} {\bibfnamefont {William~D.}\
  \bibnamefont {Pi\~neros}}\ and\ \bibinfo {author} {\bibfnamefont {\'Etienne}\
  \bibnamefont {Fodor}},\ }\bibfield  {title} {\enquote {\bibinfo {title}
  {Biased ensembles of pulsating active matter},}\ }\href {\doibase
  10.1103/PhysRevLett.134.038301} {\bibfield  {journal} {\bibinfo  {journal}
  {Phys. Rev. Lett.}\ }\textbf {\bibinfo {volume} {134}},\ \bibinfo {pages}
  {038301} (\bibinfo {year} {2025})}\BibitemShut {NoStop}%
\bibitem [{\citenamefont {Manacorda}\ and\ \citenamefont
  {Fodor}(2025)}]{manacorda2023pulsating}%
  \BibitemOpen
  \bibfield  {author} {\bibinfo {author} {\bibfnamefont {Alessandro}\
  \bibnamefont {Manacorda}}\ and\ \bibinfo {author} {\bibfnamefont {\'Etienne}\
  \bibnamefont {Fodor}},\ }\bibfield  {title} {\enquote {\bibinfo {title}
  {Diffusive oscillators capture the pulsating states of deformable
  particles},}\ }\href {\doibase 10.1103/PhysRevE.111.L053401} {\bibfield
  {journal} {\bibinfo  {journal} {Phys. Rev. E}\ }\textbf {\bibinfo {volume}
  {111}},\ \bibinfo {pages} {L053401} (\bibinfo {year} {2025})}\BibitemShut
  {NoStop}%
\bibitem [{\citenamefont {Casagrande}\ \emph
  {et~al.}(2026{\natexlab{a}})\citenamefont {Casagrande}, \citenamefont
  {Manacorda},\ and\ \citenamefont {Étienne Fodor}}]{casagrande2026a}%
  \BibitemOpen
  \bibfield  {author} {\bibinfo {author} {\bibfnamefont {Luca}\ \bibnamefont
  {Casagrande}}, \bibinfo {author} {\bibfnamefont {Alessandro}\ \bibnamefont
  {Manacorda}}, \ and\ \bibinfo {author} {\bibnamefont {Étienne Fodor}},\
  }\href {https://arxiv.org/abs/2605.25996} {\enquote {\bibinfo {title}
  {Topology of pulsating active matter: Defect asymmetry controls emergent
  motility},}\ } (\bibinfo {year} {2026}{\natexlab{a}}),\ \Eprint
  {http://arxiv.org/abs/2605.25996} {arXiv:2605.25996 [cond-mat.soft]}
  \BibitemShut {NoStop}%
\bibitem [{\citenamefont {Casagrande}\ \emph
  {et~al.}(2026{\natexlab{b}})\citenamefont {Casagrande}, \citenamefont
  {Manacorda},\ and\ \citenamefont {Fodor}}]{casagrande2026b}%
  \BibitemOpen
  \bibfield  {author} {\bibinfo {author} {\bibfnamefont {Luca}\ \bibnamefont
  {Casagrande}}, \bibinfo {author} {\bibfnamefont {Alessandro}\ \bibnamefont
  {Manacorda}}, \ and\ \bibinfo {author} {\bibfnamefont {Etienne}\ \bibnamefont
  {Fodor}},\ }\href {https://arxiv.org/abs/2605.25744} {\enquote {\bibinfo
  {title} {Collective deformation of anisotropic particles with internal
  pulsation},}\ } (\bibinfo {year} {2026}{\natexlab{b}}),\ \Eprint
  {http://arxiv.org/abs/2605.25744} {arXiv:2605.25744 [cond-mat.soft]}
  \BibitemShut {NoStop}%
\bibitem [{\citenamefont {Aranson}\ and\ \citenamefont
  {Kramer}(2002{\natexlab{a}})}]{aranson-review}%
  \BibitemOpen
  \bibfield  {author} {\bibinfo {author} {\bibfnamefont {Igor~S.}\ \bibnamefont
  {Aranson}}\ and\ \bibinfo {author} {\bibfnamefont {Lorenz}\ \bibnamefont
  {Kramer}},\ }\bibfield  {title} {\enquote {\bibinfo {title} {The world of the
  complex ginzburg-landau equation},}\ }\href {\doibase
  10.1103/RevModPhys.74.99} {\bibfield  {journal} {\bibinfo  {journal} {Rev.
  Mod. Phys.}\ }\textbf {\bibinfo {volume} {74}},\ \bibinfo {pages} {99--143}
  (\bibinfo {year} {2002}{\natexlab{a}})}\BibitemShut {NoStop}%
\bibitem [{\citenamefont {{See Supplemental Material at [url to be inserted by
  publisher] for videos corresponding to Figs.~1-3}}()}]{Movie_defects}%
  \BibitemOpen
  \bibfield  {author} {\bibinfo {author} {\bibnamefont {{See Supplemental
  Material at [url to be inserted by publisher] for videos corresponding to
  Figs.~1-3}}},\ }\href@noop {} {}\BibitemShut {NoStop}%
\bibitem [{\citenamefont {{All necessary data supporting the main findings are
  available in the paper. Necessary codes can be found at the following
  URL}}()}]{data}%
  \BibitemOpen
  \bibfield  {author} {\bibinfo {author} {\bibnamefont {{All necessary data
  supporting the main findings are available in the paper. Necessary codes can
  be found at the following URL}}},\ }\href@noop {} {}\bibinfo {howpublished}
  {\url{https://github.com/tirthankar698/Hydrodynamics-of-pulsating-matter/blob/main/Pulsating_hydro.zip}}\BibitemShut
  {NoStop}%
\bibitem [{\citenamefont {Zehnder}\ \emph {et~al.}(2015)\citenamefont
  {Zehnder}, \citenamefont {Suaris}, \citenamefont {Bellaire},\ and\
  \citenamefont {Angelini}}]{Zehnder2015}%
  \BibitemOpen
  \bibfield  {author} {\bibinfo {author} {\bibfnamefont {Steven~M.}\
  \bibnamefont {Zehnder}}, \bibinfo {author} {\bibfnamefont {Melanie}\
  \bibnamefont {Suaris}}, \bibinfo {author} {\bibfnamefont {Madisonclaire~M.}\
  \bibnamefont {Bellaire}}, \ and\ \bibinfo {author} {\bibfnamefont
  {Thomas~E.}\ \bibnamefont {Angelini}},\ }\bibfield  {title} {\enquote
  {\bibinfo {title} {Cell volume fluctuations in mdck monolayers},}\ }\href
  {\doibase 10.1016/j.bpj.2014.11.1856} {\bibfield  {journal} {\bibinfo
  {journal} {Biophys. J.}\ }\textbf {\bibinfo {volume} {108}},\ \bibinfo
  {pages} {247–250} (\bibinfo {year} {2015})}\BibitemShut {NoStop}%
\bibitem [{\citenamefont {Thiagarajan}\ \emph {et~al.}(2022)\citenamefont
  {Thiagarajan}, \citenamefont {Bhat}, \citenamefont {Salbreux}, \citenamefont
  {Inamdar},\ and\ \citenamefont {Riveline}}]{THIAGARAJAN2022105053}%
  \BibitemOpen
  \bibfield  {author} {\bibinfo {author} {\bibfnamefont {Raghavan}\
  \bibnamefont {Thiagarajan}}, \bibinfo {author} {\bibfnamefont {Alka}\
  \bibnamefont {Bhat}}, \bibinfo {author} {\bibfnamefont {Guillaume}\
  \bibnamefont {Salbreux}}, \bibinfo {author} {\bibfnamefont {Mandar~M.}\
  \bibnamefont {Inamdar}}, \ and\ \bibinfo {author} {\bibfnamefont {Daniel}\
  \bibnamefont {Riveline}},\ }\bibfield  {title} {\enquote {\bibinfo {title}
  {Pulsations and flows in tissues as two collective dynamics with simple
  cellular rules},}\ }\href {\doibase
  https://doi.org/10.1016/j.isci.2022.105053} {\bibfield  {journal} {\bibinfo
  {journal} {iScience}\ }\textbf {\bibinfo {volume} {25}},\ \bibinfo {pages}
  {105053} (\bibinfo {year} {2022})}\BibitemShut {NoStop}%
\bibitem [{\citenamefont {Angelini}\ \emph {et~al.}(2011)\citenamefont
  {Angelini}, \citenamefont {Hannezo}, \citenamefont {Trepat}, \citenamefont
  {Marquez}, \citenamefont {Fredberg},\ and\ \citenamefont
  {Weitz}}]{Hannezo-glassy-tissue}%
  \BibitemOpen
  \bibfield  {author} {\bibinfo {author} {\bibfnamefont {Thomas~E.}\
  \bibnamefont {Angelini}}, \bibinfo {author} {\bibfnamefont {Edouard}\
  \bibnamefont {Hannezo}}, \bibinfo {author} {\bibfnamefont {Xavier}\
  \bibnamefont {Trepat}}, \bibinfo {author} {\bibfnamefont {Manuel}\
  \bibnamefont {Marquez}}, \bibinfo {author} {\bibfnamefont {Jeffrey~J.}\
  \bibnamefont {Fredberg}}, \ and\ \bibinfo {author} {\bibfnamefont {David~A.}\
  \bibnamefont {Weitz}},\ }\bibfield  {title} {\enquote {\bibinfo {title}
  {Glass-like dynamics of collective cell migration},}\ }\href {\doibase
  10.1073/pnas.1010059108} {\bibfield  {journal} {\bibinfo  {journal} {Proc.
  Natl. Acad. Sci. U.S.A}\ }\textbf {\bibinfo {volume} {108}},\ \bibinfo
  {pages} {4714--4719} (\bibinfo {year} {2011})}\BibitemShut {NoStop}%
\bibitem [{\citenamefont {Banerjee}\ \emph {et~al.}(2024)\citenamefont
  {Banerjee}, \citenamefont {Desaleux}, \citenamefont {Ranft},\ and\
  \citenamefont {Étienne Fodor}}]{ALP}%
  \BibitemOpen
  \bibfield  {author} {\bibinfo {author} {\bibfnamefont {Tirthankar}\
  \bibnamefont {Banerjee}}, \bibinfo {author} {\bibfnamefont {Thibault}\
  \bibnamefont {Desaleux}}, \bibinfo {author} {\bibfnamefont {Jonas}\
  \bibnamefont {Ranft}}, \ and\ \bibinfo {author} {\bibnamefont {Étienne
  Fodor}},\ }\href {https://arxiv.org/abs/2407.19955} {\enquote {\bibinfo
  {title} {Hydrodynamics of pulsating active liquids},}\ } (\bibinfo {year}
  {2024}),\ \Eprint {http://arxiv.org/abs/2407.19955} {arXiv:2407.19955
  [cond-mat.soft]} \BibitemShut {NoStop}%
\bibitem [{\citenamefont {Aranson}\ and\ \citenamefont
  {Kramer}(2002{\natexlab{b}})}]{Aranson2002}%
  \BibitemOpen
  \bibfield  {author} {\bibinfo {author} {\bibfnamefont {Igor~S.}\ \bibnamefont
  {Aranson}}\ and\ \bibinfo {author} {\bibfnamefont {Lorenz}\ \bibnamefont
  {Kramer}},\ }\bibfield  {title} {\enquote {\bibinfo {title} {The world of the
  complex ginzburg-landau equation},}\ }\href {\doibase
  10.1103/RevModPhys.74.99} {\bibfield  {journal} {\bibinfo  {journal} {Rev.
  Mod. Phys.}\ }\textbf {\bibinfo {volume} {74}},\ \bibinfo {pages} {99--143}
  (\bibinfo {year} {2002}{\natexlab{b}})}\BibitemShut {NoStop}%
\bibitem [{\citenamefont {Lindner}\ \emph {et~al.}(2004)\citenamefont
  {Lindner}, \citenamefont {Garc\'ia-Ojalvo}, \citenamefont {Neiman},\ and\
  \citenamefont {Schimansky-Geier}}]{LINDNER2004321}%
  \BibitemOpen
  \bibfield  {author} {\bibinfo {author} {\bibfnamefont {B.}~\bibnamefont
  {Lindner}}, \bibinfo {author} {\bibfnamefont {J.}~\bibnamefont
  {Garc\'ia-Ojalvo}}, \bibinfo {author} {\bibfnamefont {A.}~\bibnamefont
  {Neiman}}, \ and\ \bibinfo {author} {\bibfnamefont {L.}~\bibnamefont
  {Schimansky-Geier}},\ }\bibfield  {title} {\enquote {\bibinfo {title}
  {Effects of noise in excitable systems},}\ }\href {\doibase
  https://doi.org/10.1016/j.physrep.2003.10.015} {\bibfield  {journal}
  {\bibinfo  {journal} {Phys. Rep.}\ }\textbf {\bibinfo {volume} {392}},\
  \bibinfo {pages} {321--424} (\bibinfo {year} {2004})}\BibitemShut {NoStop}%
\bibitem [{\citenamefont {Aliev}\ and\ \citenamefont
  {Panfilov}(1996)}]{ALIEV1996293}%
  \BibitemOpen
  \bibfield  {author} {\bibinfo {author} {\bibfnamefont {Rubin~R.}\
  \bibnamefont {Aliev}}\ and\ \bibinfo {author} {\bibfnamefont {Alexander~V.}\
  \bibnamefont {Panfilov}},\ }\bibfield  {title} {\enquote {\bibinfo {title} {A
  simple two-variable model of cardiac excitation},}\ }\href {\doibase
  https://doi.org/10.1016/0960-0779(95)00089-5} {\bibfield  {journal} {\bibinfo
   {journal} {Chaos, Solitons and Fractals}\ }\textbf {\bibinfo {volume} {7}},\
  \bibinfo {pages} {293--301} (\bibinfo {year} {1996})}\BibitemShut {NoStop}%
\bibitem [{\citenamefont {Gani}\ and\ \citenamefont
  {Ogawa}(2016)}]{GANI201630}%
  \BibitemOpen
  \bibfield  {author} {\bibinfo {author} {\bibfnamefont {M.~Osman}\
  \bibnamefont {Gani}}\ and\ \bibinfo {author} {\bibfnamefont {Toshiyuki}\
  \bibnamefont {Ogawa}},\ }\bibfield  {title} {\enquote {\bibinfo {title}
  {Stability of periodic traveling waves in the aliev–panfilov
  reaction–diffusion system},}\ }\href {\doibase
  https://doi.org/10.1016/j.cnsns.2015.09.002} {\bibfield  {journal} {\bibinfo
  {journal} {Commun. Nonlinear Sci. Numer. Simul.}\ }\textbf {\bibinfo {volume}
  {33}},\ \bibinfo {pages} {30--42} (\bibinfo {year} {2016})}\BibitemShut
  {NoStop}%
\bibitem [{\citenamefont {Cebrián-Lacasa}\ \emph {et~al.}(2024)\citenamefont
  {Cebrián-Lacasa}, \citenamefont {Parra-Rivas}, \citenamefont
  {Ruiz-Reynés},\ and\ \citenamefont {Gelens}}]{FN-review}%
  \BibitemOpen
  \bibfield  {author} {\bibinfo {author} {\bibfnamefont {Daniel}\ \bibnamefont
  {Cebrián-Lacasa}}, \bibinfo {author} {\bibfnamefont {Pedro}\ \bibnamefont
  {Parra-Rivas}}, \bibinfo {author} {\bibfnamefont {Daniel}\ \bibnamefont
  {Ruiz-Reynés}}, \ and\ \bibinfo {author} {\bibfnamefont {Lendert}\
  \bibnamefont {Gelens}},\ }\bibfield  {title} {\enquote {\bibinfo {title} {Six
  decades of the fitzhugh–nagumo model: A guide through its spatio-temporal
  dynamics and influence across disciplines},}\ }\href {\doibase
  https://doi.org/10.1016/j.physrep.2024.09.014} {\bibfield  {journal}
  {\bibinfo  {journal} {Phys. Rep.}\ }\textbf {\bibinfo {volume} {1096}},\
  \bibinfo {pages} {1--39} (\bibinfo {year} {2024})}\BibitemShut {NoStop}%
\bibitem [{\citenamefont {Sakaguchi}\ and\ \citenamefont
  {Maeyama}(2014)}]{Sakaguchi-Maeyama}%
  \BibitemOpen
  \bibfield  {author} {\bibinfo {author} {\bibfnamefont {Hidetsugu}\
  \bibnamefont {Sakaguchi}}\ and\ \bibinfo {author} {\bibfnamefont {Satomi}\
  \bibnamefont {Maeyama}},\ }\bibfield  {title} {\enquote {\bibinfo {title}
  {Competitive aggregation dynamics using phase wave signals},}\ }\href
  {\doibase 10.1016/j.jtbi.2014.06.017} {\bibfield  {journal} {\bibinfo
  {journal} {J. Theor. Biol.}\ }\textbf {\bibinfo {volume} {359}},\ \bibinfo
  {pages} {155–160} (\bibinfo {year} {2014})}\BibitemShut {NoStop}%
\bibitem [{\citenamefont {Banerjee}\ and\ \citenamefont
  {Basu}(2017)}]{tirtha-oscillator}%
  \BibitemOpen
  \bibfield  {author} {\bibinfo {author} {\bibfnamefont {Tirthankar}\
  \bibnamefont {Banerjee}}\ and\ \bibinfo {author} {\bibfnamefont {Abhik}\
  \bibnamefont {Basu}},\ }\bibfield  {title} {\enquote {\bibinfo {title}
  {Active hydrodynamics of synchronization and ordering in moving
  oscillators},}\ }\href {\doibase 10.1103/PhysRevE.96.022201} {\bibfield
  {journal} {\bibinfo  {journal} {Phys. Rev. E}\ }\textbf {\bibinfo {volume}
  {96}},\ \bibinfo {pages} {022201} (\bibinfo {year} {2017})}\BibitemShut
  {NoStop}%
\bibitem [{\citenamefont {Cross}\ and\ \citenamefont
  {Greenside}(2009)}]{Cross_Greenside_2009}%
  \BibitemOpen
  \bibfield  {author} {\bibinfo {author} {\bibfnamefont {Michael}\ \bibnamefont
  {Cross}}\ and\ \bibinfo {author} {\bibfnamefont {Henry}\ \bibnamefont
  {Greenside}},\ }\href {\doibase https://doi.org/10.1017/CBO9780511627200}
  {\emph {\bibinfo {title} {Pattern Formation and Dynamics in Nonequilibrium
  Systems}}}\ (\bibinfo  {publisher} {Cambridge University Press, Cambridge},\
  \bibinfo {year} {2009})\BibitemShut {NoStop}%
\bibitem [{\citenamefont {Cross}\ and\ \citenamefont
  {Hohenberg}(1993)}]{hohenberg-review}%
  \BibitemOpen
  \bibfield  {author} {\bibinfo {author} {\bibfnamefont {M.~C.}\ \bibnamefont
  {Cross}}\ and\ \bibinfo {author} {\bibfnamefont {P.~C.}\ \bibnamefont
  {Hohenberg}},\ }\bibfield  {title} {\enquote {\bibinfo {title} {Pattern
  formation outside of equilibrium},}\ }\href {\doibase
  10.1103/RevModPhys.65.851} {\bibfield  {journal} {\bibinfo  {journal} {Rev.
  Mod. Phys.}\ }\textbf {\bibinfo {volume} {65}},\ \bibinfo {pages} {851--1112}
  (\bibinfo {year} {1993})}\BibitemShut {NoStop}%
\bibitem [{\citenamefont {Mazenko}(1997)}]{Mazenko1997}%
  \BibitemOpen
  \bibfield  {author} {\bibinfo {author} {\bibfnamefont {Gene~F.}\ \bibnamefont
  {Mazenko}},\ }\bibfield  {title} {\enquote {\bibinfo {title} {Vortex
  velocities in the $\mathit{O}(\mathit{n})$ symmetric time-dependent
  ginzburg-landau model},}\ }\href {\doibase 10.1103/PhysRevLett.78.401}
  {\bibfield  {journal} {\bibinfo  {journal} {Phys. Rev. Lett.}\ }\textbf
  {\bibinfo {volume} {78}},\ \bibinfo {pages} {401--404} (\bibinfo {year}
  {1997})}\BibitemShut {NoStop}%
\bibitem [{\citenamefont {Qi}\ and\ \citenamefont {Chen}(2008)}]{qi2008}%
  \BibitemOpen
  \bibfield  {author} {\bibinfo {author} {\bibfnamefont {Wei-Kai}\ \bibnamefont
  {Qi}}\ and\ \bibinfo {author} {\bibfnamefont {Yong}\ \bibnamefont {Chen}},\
  }\href {https://arxiv.org/abs/0809.0348} {\enquote {\bibinfo {title}
  {Topological dynamics and dynamical scaling behavior of vortices in a
  two-dimensional xy model},}\ } (\bibinfo {year} {2008}),\ \Eprint
  {http://arxiv.org/abs/0809.0348} {arXiv:0809.0348 [cond-mat.stat-mech]}
  \BibitemShut {NoStop}%
\bibitem [{\citenamefont {Skogvoll}\ \emph {et~al.}(2023)\citenamefont
  {Skogvoll}, \citenamefont {R{\o}nning}, \citenamefont {Salvalaglio},\ and\
  \citenamefont {Angheluta}}]{Skogvoll2023}%
  \BibitemOpen
  \bibfield  {author} {\bibinfo {author} {\bibfnamefont {Vidar}\ \bibnamefont
  {Skogvoll}}, \bibinfo {author} {\bibfnamefont {Jonas}\ \bibnamefont
  {R{\o}nning}}, \bibinfo {author} {\bibfnamefont {Marco}\ \bibnamefont
  {Salvalaglio}}, \ and\ \bibinfo {author} {\bibfnamefont {Luiza}\ \bibnamefont
  {Angheluta}},\ }\bibfield  {title} {\enquote {\bibinfo {title} {A unified
  field theory of topological defects and non-linear local excitations},}\
  }\href {\doibase 10.1038/s41524-023-01077-6} {\bibfield  {journal} {\bibinfo
  {journal} {npj Computational Materials}\ }\textbf {\bibinfo {volume} {9}},\
  \bibinfo {pages} {122} (\bibinfo {year} {2023})}\BibitemShut {NoStop}%
\bibitem [{\citenamefont {Angheluta}\ \emph {et~al.}(2021)\citenamefont
  {Angheluta}, \citenamefont {Chen}, \citenamefont {Marchetti},\ and\
  \citenamefont {Bowick}}]{Angheluta_2021}%
  \BibitemOpen
  \bibfield  {author} {\bibinfo {author} {\bibfnamefont {Luiza}\ \bibnamefont
  {Angheluta}}, \bibinfo {author} {\bibfnamefont {Zhitao}\ \bibnamefont
  {Chen}}, \bibinfo {author} {\bibfnamefont {M~Cristina}\ \bibnamefont
  {Marchetti}}, \ and\ \bibinfo {author} {\bibfnamefont {Mark~J}\ \bibnamefont
  {Bowick}},\ }\bibfield  {title} {\enquote {\bibinfo {title} {The role of
  fluid flow in the dynamics of active nematic defects},}\ }\href {\doibase
  10.1088/1367-2630/abe8a8} {\bibfield  {journal} {\bibinfo  {journal} {New
  Journal of Physics}\ }\textbf {\bibinfo {volume} {23}},\ \bibinfo {pages}
  {033009} (\bibinfo {year} {2021})}\BibitemShut {NoStop}%
\bibitem [{\citenamefont {Angheluta}\ \emph {et~al.}(2012)\citenamefont
  {Angheluta}, \citenamefont {Jeraldo},\ and\ \citenamefont
  {Goldenfeld}}]{PhysRevE.85.011153}%
  \BibitemOpen
  \bibfield  {author} {\bibinfo {author} {\bibfnamefont {Luiza}\ \bibnamefont
  {Angheluta}}, \bibinfo {author} {\bibfnamefont {Patricio}\ \bibnamefont
  {Jeraldo}}, \ and\ \bibinfo {author} {\bibfnamefont {Nigel}\ \bibnamefont
  {Goldenfeld}},\ }\bibfield  {title} {\enquote {\bibinfo {title} {Anisotropic
  velocity statistics of topological defects under shear flow},}\ }\href
  {\doibase 10.1103/PhysRevE.85.011153} {\bibfield  {journal} {\bibinfo
  {journal} {Phys. Rev. E}\ }\textbf {\bibinfo {volume} {85}},\ \bibinfo
  {pages} {011153} (\bibinfo {year} {2012})}\BibitemShut {NoStop}%
\bibitem [{\citenamefont {Bray}(1994)}]{Bray1994}%
  \BibitemOpen
  \bibfield  {author} {\bibinfo {author} {\bibfnamefont {A.~J.}\ \bibnamefont
  {Bray}},\ }\bibfield  {title} {\enquote {\bibinfo {title} {Theory of
  phase-ordering kinetics},}\ }\href {\doibase 10.1080/00018739400101505}
  {\bibfield  {journal} {\bibinfo  {journal} {Adv. Phys.}\ }\textbf {\bibinfo
  {volume} {43}},\ \bibinfo {pages} {357--459} (\bibinfo {year}
  {1994})}\BibitemShut {NoStop}%
\bibitem [{\citenamefont {Maryshev}\ \emph {et~al.}(2020)\citenamefont
  {Maryshev}, \citenamefont {Morozov}, \citenamefont {Goryachev},\ and\
  \citenamefont {Marenduzzo}}]{Maryshev-actmodelC}%
  \BibitemOpen
  \bibfield  {author} {\bibinfo {author} {\bibfnamefont {Ivan}\ \bibnamefont
  {Maryshev}}, \bibinfo {author} {\bibfnamefont {Alexander}\ \bibnamefont
  {Morozov}}, \bibinfo {author} {\bibfnamefont {Andrew~B.}\ \bibnamefont
  {Goryachev}}, \ and\ \bibinfo {author} {\bibfnamefont {Davide}\ \bibnamefont
  {Marenduzzo}},\ }\bibfield  {title} {\enquote {\bibinfo {title} {Pattern
  formation in active model c with anchoring: bands{,} aster networks{,} and
  foams},}\ }\href {\doibase 10.1039/D0SM00927J} {\bibfield  {journal}
  {\bibinfo  {journal} {Soft Matter}\ }\textbf {\bibinfo {volume} {16}},\
  \bibinfo {pages} {8775--8781} (\bibinfo {year} {2020})}\BibitemShut {NoStop}%
\bibitem [{\citenamefont {Ra\ss{}hofer}\ \emph {et~al.}(2025)\citenamefont
  {Ra\ss{}hofer}, \citenamefont {Bauer}, \citenamefont {Ziepke}, \citenamefont
  {Maryshev},\ and\ \citenamefont {Frey}}]{Erwin-actmodelC}%
  \BibitemOpen
  \bibfield  {author} {\bibinfo {author} {\bibfnamefont {Florian}\ \bibnamefont
  {Ra\ss{}hofer}}, \bibinfo {author} {\bibfnamefont {Simon}\ \bibnamefont
  {Bauer}}, \bibinfo {author} {\bibfnamefont {Alexander}\ \bibnamefont
  {Ziepke}}, \bibinfo {author} {\bibfnamefont {Ivan}\ \bibnamefont {Maryshev}},
  \ and\ \bibinfo {author} {\bibfnamefont {Erwin}\ \bibnamefont {Frey}},\
  }\bibfield  {title} {\enquote {\bibinfo {title} {Capillary wave formation in
  conserved active emulsions},}\ }\href {\doibase 10.1103/lqws-dkmy} {\bibfield
   {journal} {\bibinfo  {journal} {Phys. Rev. Res.}\ }\textbf {\bibinfo
  {volume} {7}},\ \bibinfo {pages} {043240} (\bibinfo {year}
  {2025})}\BibitemShut {NoStop}%
\bibitem [{\citenamefont {Dullweber}\ \emph {et~al.}(2025)\citenamefont
  {Dullweber}, \citenamefont {Belousov},\ and\ \citenamefont
  {Erzberger}}]{mmz3-kbrv}%
  \BibitemOpen
  \bibfield  {author} {\bibinfo {author} {\bibfnamefont {Tim}\ \bibnamefont
  {Dullweber}}, \bibinfo {author} {\bibfnamefont {Roman}\ \bibnamefont
  {Belousov}}, \ and\ \bibinfo {author} {\bibfnamefont {Anna}\ \bibnamefont
  {Erzberger}},\ }\bibfield  {title} {\enquote {\bibinfo {title} {Feedback
  between microscopic activity and macroscopic dynamics drives excitability and
  oscillations in mechanochemical matter},}\ }\href {\doibase
  10.1103/mmz3-kbrv} {\bibfield  {journal} {\bibinfo  {journal} {Phys. Rev. E}\
  }\textbf {\bibinfo {volume} {112}},\ \bibinfo {pages} {034411} (\bibinfo
  {year} {2025})}\BibitemShut {NoStop}%
\bibitem [{\citenamefont {Cocconi}\ \emph {et~al.}(2025)\citenamefont
  {Cocconi}, \citenamefont {Chatzittofi},\ and\ \citenamefont
  {Golestanian}}]{cocconi2025}%
  \BibitemOpen
  \bibfield  {author} {\bibinfo {author} {\bibfnamefont {Luca}\ \bibnamefont
  {Cocconi}}, \bibinfo {author} {\bibfnamefont {Michalis}\ \bibnamefont
  {Chatzittofi}}, \ and\ \bibinfo {author} {\bibfnamefont {Ramin}\ \bibnamefont
  {Golestanian}},\ }\bibfield  {title} {\enquote {\bibinfo {title} {Mechanical
  inhibition of dissipation in a thermodynamically consistent active solid},}\
  }\href {\doibase 10.1103/tgyv-5vqv} {\bibfield  {journal} {\bibinfo
  {journal} {Phys. Rev. Res.}\ }\textbf {\bibinfo {volume} {7}},\ \bibinfo
  {pages} {L042062} (\bibinfo {year} {2025})}\BibitemShut {NoStop}%
\bibitem [{\citenamefont {Thampi}\ \emph {et~al.}(2013)\citenamefont {Thampi},
  \citenamefont {Ansumali}, \citenamefont {Adhikari},\ and\ \citenamefont
  {Succi}}]{thampi-ronojoy}%
  \BibitemOpen
  \bibfield  {author} {\bibinfo {author} {\bibfnamefont {Sumesh~P.}\
  \bibnamefont {Thampi}}, \bibinfo {author} {\bibfnamefont {Santosh}\
  \bibnamefont {Ansumali}}, \bibinfo {author} {\bibfnamefont {R.}~\bibnamefont
  {Adhikari}}, \ and\ \bibinfo {author} {\bibfnamefont {Sauro}\ \bibnamefont
  {Succi}},\ }\bibfield  {title} {\enquote {\bibinfo {title} {Isotropic
  discrete laplacian operators from lattice hydrodynamics},}\ }\href {\doibase
  https://doi.org/10.1016/j.jcp.2012.07.037} {\bibfield  {journal} {\bibinfo
  {journal} {J. Comput. Phys.}\ }\textbf {\bibinfo {volume} {234}},\ \bibinfo
  {pages} {1--7} (\bibinfo {year} {2013})}\BibitemShut {NoStop}%
\bibitem [{\citenamefont {Milnor}(1969)}]{morse}%
  \BibitemOpen
  \bibfield  {author} {\bibinfo {author} {\bibfnamefont {J.}~\bibnamefont
  {Milnor}},\ }\href@noop {} {\emph {\bibinfo {title} {{Morse Theory}}}}\
  (\bibinfo  {publisher} {Princeton University Press},\ \bibinfo {year}
  {1969})\BibitemShut {NoStop}%
\bibitem [{\citenamefont {Chaikin}\ and\ \citenamefont
  {Lubensky}(1995)}]{Chaikin}%
  \BibitemOpen
  \bibfield  {author} {\bibinfo {author} {\bibfnamefont {Paul~M.}\ \bibnamefont
  {Chaikin}}\ and\ \bibinfo {author} {\bibfnamefont {Tom~C.}\ \bibnamefont
  {Lubensky}},\ }\href@noop {} {\emph {\bibinfo {title} {{Principles of
  condensed matter physics}}}}\ (\bibinfo  {publisher} {Cambridge University
  Press},\ \bibinfo {year} {1995})\BibitemShut {NoStop}%
\bibitem [{\citenamefont {Angheluta}\ \emph {et~al.}(2026)\citenamefont
  {Angheluta}, \citenamefont {Lång}, \citenamefont {Lång},\ and\
  \citenamefont {Bøe}}]{Angheluta2026}%
  \BibitemOpen
  \bibfield  {author} {\bibinfo {author} {\bibfnamefont {Luiza}\ \bibnamefont
  {Angheluta}}, \bibinfo {author} {\bibfnamefont {Anna}\ \bibnamefont {Lång}},
  \bibinfo {author} {\bibfnamefont {Emma}\ \bibnamefont {Lång}}, \ and\
  \bibinfo {author} {\bibfnamefont {Stig~Ove}\ \bibnamefont {Bøe}},\
  }\bibfield  {title} {\enquote {\bibinfo {title} {Full-integer topological
  defects in polar active matter: From collective migration to tissue
  patterning},}\ }\href {\doibase
  https://doi.org/10.1146/annurev-conmatphys-031620-105420} {\bibfield
  {journal} {\bibinfo  {journal} {Annu. Rev. Condens. Matter Phys.}\ }\textbf
  {\bibinfo {volume} {17}},\ \bibinfo {pages} {305--325} (\bibinfo {year}
  {2026})}\BibitemShut {NoStop}%
\bibitem [{\citenamefont {Tan}\ \emph {et~al.}(2020)\citenamefont {Tan},
  \citenamefont {Liu}, \citenamefont {Miller}, \citenamefont {Tekant},
  \citenamefont {Dunkel},\ and\ \citenamefont {Fakhri}}]{Tan2020}%
  \BibitemOpen
  \bibfield  {author} {\bibinfo {author} {\bibfnamefont {Tzer~Han}\
  \bibnamefont {Tan}}, \bibinfo {author} {\bibfnamefont {Jinghui}\ \bibnamefont
  {Liu}}, \bibinfo {author} {\bibfnamefont {Pearson~W.}\ \bibnamefont
  {Miller}}, \bibinfo {author} {\bibfnamefont {Melis}\ \bibnamefont {Tekant}},
  \bibinfo {author} {\bibfnamefont {J{\"o}rn}\ \bibnamefont {Dunkel}}, \ and\
  \bibinfo {author} {\bibfnamefont {Nikta}\ \bibnamefont {Fakhri}},\ }\bibfield
   {title} {\enquote {\bibinfo {title} {Topological turbulence in the membrane
  of a living cell},}\ }\href {\doibase 10.1038/s41567-020-0841-9} {\bibfield
  {journal} {\bibinfo  {journal} {Nat. Phys.}\ }\textbf {\bibinfo {volume}
  {16}},\ \bibinfo {pages} {657--662} (\bibinfo {year} {2020})}\BibitemShut
  {NoStop}%
\bibitem [{\citenamefont {Toner}(2012)}]{Toner2012}%
  \BibitemOpen
  \bibfield  {author} {\bibinfo {author} {\bibfnamefont {John}\ \bibnamefont
  {Toner}},\ }\bibfield  {title} {\enquote {\bibinfo {title} {Birth, death, and
  flight: A theory of malthusian flocks},}\ }\href {\doibase
  10.1103/PhysRevLett.108.088102} {\bibfield  {journal} {\bibinfo  {journal}
  {Phys. Rev. Lett.}\ }\textbf {\bibinfo {volume} {108}},\ \bibinfo {pages}
  {088102} (\bibinfo {year} {2012})}\BibitemShut {NoStop}%
\bibitem [{\citenamefont {Besse}\ \emph {et~al.}(2022)\citenamefont {Besse},
  \citenamefont {Chat\'e},\ and\ \citenamefont {Solon}}]{Besse2022}%
  \BibitemOpen
  \bibfield  {author} {\bibinfo {author} {\bibfnamefont {Marc}\ \bibnamefont
  {Besse}}, \bibinfo {author} {\bibfnamefont {Hugues}\ \bibnamefont {Chat\'e}},
  \ and\ \bibinfo {author} {\bibfnamefont {Alexandre}\ \bibnamefont {Solon}},\
  }\bibfield  {title} {\enquote {\bibinfo {title} {Metastability of
  constant-density flocks},}\ }\href {\doibase 10.1103/PhysRevLett.129.268003}
  {\bibfield  {journal} {\bibinfo  {journal} {Phys. Rev. Lett.}\ }\textbf
  {\bibinfo {volume} {129}},\ \bibinfo {pages} {268003} (\bibinfo {year}
  {2022})}\BibitemShut {NoStop}%
\bibitem [{\citenamefont {Dadhichi}\ \emph {et~al.}(2020)\citenamefont
  {Dadhichi}, \citenamefont {Kethapelli}, \citenamefont {Chajwa}, \citenamefont
  {Ramaswamy},\ and\ \citenamefont {Maitra}}]{Dadhichi2020}%
  \BibitemOpen
  \bibfield  {author} {\bibinfo {author} {\bibfnamefont {Lokrshi~Prawar}\
  \bibnamefont {Dadhichi}}, \bibinfo {author} {\bibfnamefont {Jitendra}\
  \bibnamefont {Kethapelli}}, \bibinfo {author} {\bibfnamefont {Rahul}\
  \bibnamefont {Chajwa}}, \bibinfo {author} {\bibfnamefont {Sriram}\
  \bibnamefont {Ramaswamy}}, \ and\ \bibinfo {author} {\bibfnamefont {Ananyo}\
  \bibnamefont {Maitra}},\ }\bibfield  {title} {\enquote {\bibinfo {title}
  {Nonmutual torques and the unimportance of motility for long-range order in
  two-dimensional flocks},}\ }\href {\doibase 10.1103/PhysRevE.101.052601}
  {\bibfield  {journal} {\bibinfo  {journal} {Phys. Rev. E}\ }\textbf {\bibinfo
  {volume} {101}},\ \bibinfo {pages} {052601} (\bibinfo {year}
  {2020})}\BibitemShut {NoStop}%
\bibitem [{\citenamefont {Dopierala}\ \emph {et~al.}(2025)\citenamefont
  {Dopierala}, \citenamefont {Chat\'e}, \citenamefont {Shi},\ and\
  \citenamefont {Solon}}]{dopierala2025}%
  \BibitemOpen
  \bibfield  {author} {\bibinfo {author} {\bibfnamefont {Dawid}\ \bibnamefont
  {Dopierala}}, \bibinfo {author} {\bibfnamefont {Hugues}\ \bibnamefont
  {Chat\'e}}, \bibinfo {author} {\bibfnamefont {Xia-qing}\ \bibnamefont {Shi}},
  \ and\ \bibinfo {author} {\bibfnamefont {Alexandre}\ \bibnamefont {Solon}},\
  }\bibfield  {title} {\enquote {\bibinfo {title} {Inescapable anisotropy of
  nonreciprocal xy models},}\ }\href {\doibase 10.1103/r3dx-7lrd} {\bibfield
  {journal} {\bibinfo  {journal} {Phys. Rev. Lett.}\ }\textbf {\bibinfo
  {volume} {135}},\ \bibinfo {pages} {088302} (\bibinfo {year}
  {2025})}\BibitemShut {NoStop}%
\bibitem [{\citenamefont {Popli}\ \emph
  {et~al.}(2025{\natexlab{a}})\citenamefont {Popli}, \citenamefont {Maitra},\
  and\ \citenamefont {Ramaswamy}}]{maitra2025}%
  \BibitemOpen
  \bibfield  {author} {\bibinfo {author} {\bibfnamefont {Pankaj}\ \bibnamefont
  {Popli}}, \bibinfo {author} {\bibfnamefont {Ananyo}\ \bibnamefont {Maitra}},
  \ and\ \bibinfo {author} {\bibfnamefont {Sriram}\ \bibnamefont {Ramaswamy}},\
  }\bibfield  {title} {\enquote {\bibinfo {title} {Ordering and defect cloaking
  in nonreciprocal lattice xy models},}\ }\href {\doibase 10.1103/2yky-45sr}
  {\bibfield  {journal} {\bibinfo  {journal} {Phys. Rev. Lett.}\ }\textbf
  {\bibinfo {volume} {135}},\ \bibinfo {pages} {088303} (\bibinfo {year}
  {2025}{\natexlab{a}})}\BibitemShut {NoStop}%
\bibitem [{\citenamefont {Popli}\ \emph
  {et~al.}(2025{\natexlab{b}})\citenamefont {Popli}, \citenamefont {Maitra},\
  and\ \citenamefont {Ramaswamy}}]{popli2025}%
  \BibitemOpen
  \bibfield  {author} {\bibinfo {author} {\bibfnamefont {Pankaj}\ \bibnamefont
  {Popli}}, \bibinfo {author} {\bibfnamefont {Ananyo}\ \bibnamefont {Maitra}},
  \ and\ \bibinfo {author} {\bibfnamefont {Sriram}\ \bibnamefont {Ramaswamy}},\
  }\bibfield  {title} {\enquote {\bibinfo {title} {Ordering and defect cloaking
  in nonreciprocal lattice xy models},}\ }\href {\doibase 10.1103/2yky-45sr}
  {\bibfield  {journal} {\bibinfo  {journal} {Phys. Rev. Lett.}\ }\textbf
  {\bibinfo {volume} {135}},\ \bibinfo {pages} {088303} (\bibinfo {year}
  {2025}{\natexlab{b}})}\BibitemShut {NoStop}%
\bibitem [{\citenamefont {Sutherland}\ \emph {et~al.}(1999)\citenamefont
  {Sutherland}, \citenamefont {Dalziel}, \citenamefont {Hughes},\ and\
  \citenamefont {Linden}}]{SUTHERLAND_99}%
  \BibitemOpen
  \bibfield  {author} {\bibinfo {author} {\bibfnamefont {Bruce~R.}\
  \bibnamefont {Sutherland}}, \bibinfo {author} {\bibfnamefont {Stuart~B.}\
  \bibnamefont {Dalziel}}, \bibinfo {author} {\bibfnamefont {Graham~O.}\
  \bibnamefont {Hughes}}, \ and\ \bibinfo {author} {\bibfnamefont {P.~F.}\
  \bibnamefont {Linden}},\ }\bibfield  {title} {\enquote {\bibinfo {title}
  {Visualization and measurement of internal waves by ‘synthetic
  schlieren’. part 1. vertically oscillating cylinder},}\ }\href {\doibase
  10.1017/S0022112099005017} {\bibfield  {journal} {\bibinfo  {journal} {J.
  Fluid Mech.}\ }\textbf {\bibinfo {volume} {390}},\ \bibinfo {pages}
  {93–126} (\bibinfo {year} {1999})}\BibitemShut {NoStop}%
\bibitem [{\citenamefont {Tang}\ \emph {et~al.}(2025)\citenamefont {Tang},
  \citenamefont {Nejad}, \citenamefont {Pegoraro}, \citenamefont {Mahadevan},\
  and\ \citenamefont {Guo}}]{tang2025}%
  \BibitemOpen
  \bibfield  {author} {\bibinfo {author} {\bibfnamefont {Wenhui}\ \bibnamefont
  {Tang}}, \bibinfo {author} {\bibfnamefont {Mehrana~R.}\ \bibnamefont
  {Nejad}}, \bibinfo {author} {\bibfnamefont {Adrian~F.}\ \bibnamefont
  {Pegoraro}}, \bibinfo {author} {\bibfnamefont {L.}~\bibnamefont {Mahadevan}},
  \ and\ \bibinfo {author} {\bibfnamefont {Ming}\ \bibnamefont {Guo}},\ }\href
  {https://arxiv.org/abs/2507.16772} {\enquote {\bibinfo {title} {Collective
  synchrony in confluent, pulsatile epithelia},}\ } (\bibinfo {year} {2025}),\
  \Eprint {http://arxiv.org/abs/2507.16772} {arXiv:2507.16772 [cond-mat.soft]}
  \BibitemShut {NoStop}%
\bibitem [{\citenamefont {Shankar}\ \emph {et~al.}(2018)\citenamefont
  {Shankar}, \citenamefont {Ramaswamy}, \citenamefont {Marchetti},\ and\
  \citenamefont {Bowick}}]{Shankar2018}%
  \BibitemOpen
  \bibfield  {author} {\bibinfo {author} {\bibfnamefont {Suraj}\ \bibnamefont
  {Shankar}}, \bibinfo {author} {\bibfnamefont {Sriram}\ \bibnamefont
  {Ramaswamy}}, \bibinfo {author} {\bibfnamefont {M.~Cristina}\ \bibnamefont
  {Marchetti}}, \ and\ \bibinfo {author} {\bibfnamefont {Mark~J.}\ \bibnamefont
  {Bowick}},\ }\bibfield  {title} {\enquote {\bibinfo {title} {Defect unbinding
  in active nematics},}\ }\href {\doibase 10.1103/PhysRevLett.121.108002}
  {\bibfield  {journal} {\bibinfo  {journal} {Phys. Rev. Lett.}\ }\textbf
  {\bibinfo {volume} {121}},\ \bibinfo {pages} {108002} (\bibinfo {year}
  {2018})}\BibitemShut {NoStop}%
\bibitem [{\citenamefont {Shankar}\ and\ \citenamefont
  {Marchetti}(2019)}]{Shankar2019}%
  \BibitemOpen
  \bibfield  {author} {\bibinfo {author} {\bibfnamefont {Suraj}\ \bibnamefont
  {Shankar}}\ and\ \bibinfo {author} {\bibfnamefont {M.~Cristina}\ \bibnamefont
  {Marchetti}},\ }\bibfield  {title} {\enquote {\bibinfo {title} {Hydrodynamics
  of active defects: From order to chaos to defect ordering},}\ }\href
  {\doibase 10.1103/PhysRevX.9.041047} {\bibfield  {journal} {\bibinfo
  {journal} {Phys. Rev. X}\ }\textbf {\bibinfo {volume} {9}},\ \bibinfo {pages}
  {041047} (\bibinfo {year} {2019})}\BibitemShut {NoStop}%
\bibitem [{\citenamefont {Shankar}\ \emph
  {et~al.}(2022{\natexlab{a}})\citenamefont {Shankar}, \citenamefont {Souslov},
  \citenamefont {Bowick}, \citenamefont {Marchetti},\ and\ \citenamefont
  {Vitelli}}]{Shankar22}%
  \BibitemOpen
  \bibfield  {author} {\bibinfo {author} {\bibfnamefont {Suraj}\ \bibnamefont
  {Shankar}}, \bibinfo {author} {\bibfnamefont {Anton}\ \bibnamefont
  {Souslov}}, \bibinfo {author} {\bibfnamefont {Mark~J.}\ \bibnamefont
  {Bowick}}, \bibinfo {author} {\bibfnamefont {M.~Cristina}\ \bibnamefont
  {Marchetti}}, \ and\ \bibinfo {author} {\bibfnamefont {Vincenzo}\
  \bibnamefont {Vitelli}},\ }\bibfield  {title} {\enquote {\bibinfo {title}
  {Topological active matter},}\ }\href {\doibase 10.1038/s42254-022-00445-3}
  {\bibfield  {journal} {\bibinfo  {journal} {Nat. Rev. Phys.}\ }\textbf
  {\bibinfo {volume} {4}},\ \bibinfo {pages} {380–398} (\bibinfo {year}
  {2022}{\natexlab{a}})}\BibitemShut {NoStop}%
\bibitem [{\citenamefont {Norton}\ \emph {et~al.}(2020)\citenamefont {Norton},
  \citenamefont {Grover}, \citenamefont {Hagan},\ and\ \citenamefont
  {Fraden}}]{Norton2020}%
  \BibitemOpen
  \bibfield  {author} {\bibinfo {author} {\bibfnamefont {Michael~M.}\
  \bibnamefont {Norton}}, \bibinfo {author} {\bibfnamefont {Piyush}\
  \bibnamefont {Grover}}, \bibinfo {author} {\bibfnamefont {Michael~F.}\
  \bibnamefont {Hagan}}, \ and\ \bibinfo {author} {\bibfnamefont {Seth}\
  \bibnamefont {Fraden}},\ }\bibfield  {title} {\enquote {\bibinfo {title}
  {Optimal control of active nematics},}\ }\href {\doibase
  10.1103/PhysRevLett.125.178005} {\bibfield  {journal} {\bibinfo  {journal}
  {Phys. Rev. Lett.}\ }\textbf {\bibinfo {volume} {125}},\ \bibinfo {pages}
  {178005} (\bibinfo {year} {2020})}\BibitemShut {NoStop}%
\bibitem [{\citenamefont {Shankar}\ \emph
  {et~al.}(2022{\natexlab{b}})\citenamefont {Shankar}, \citenamefont {Raju},\
  and\ \citenamefont {Mahadevan}}]{Shankar2022}%
  \BibitemOpen
  \bibfield  {author} {\bibinfo {author} {\bibfnamefont {Suraj}\ \bibnamefont
  {Shankar}}, \bibinfo {author} {\bibfnamefont {Vidya}\ \bibnamefont {Raju}}, \
  and\ \bibinfo {author} {\bibfnamefont {L.}~\bibnamefont {Mahadevan}},\
  }\bibfield  {title} {\enquote {\bibinfo {title} {Optimal transport and
  control of active drops},}\ }\href {\doibase 10.1073/pnas.2121985119}
  {\bibfield  {journal} {\bibinfo  {journal} {Proc. Natl. Acad. Sci. USA}\
  }\textbf {\bibinfo {volume} {119}},\ \bibinfo {pages} {e2121985119} (\bibinfo
  {year} {2022}{\natexlab{b}})}\BibitemShut {NoStop}%
\bibitem [{\citenamefont {Davis}\ \emph {et~al.}(2024)\citenamefont {Davis},
  \citenamefont {Proesmans},\ and\ \citenamefont {Fodor}}]{Davis2025}%
  \BibitemOpen
  \bibfield  {author} {\bibinfo {author} {\bibfnamefont {Luke~K.}\ \bibnamefont
  {Davis}}, \bibinfo {author} {\bibfnamefont {Karel}\ \bibnamefont
  {Proesmans}}, \ and\ \bibinfo {author} {\bibfnamefont {\'Etienne}\
  \bibnamefont {Fodor}},\ }\bibfield  {title} {\enquote {\bibinfo {title}
  {Active matter under control: Insights from response theory},}\ }\href
  {\doibase 10.1103/PhysRevX.14.011012} {\bibfield  {journal} {\bibinfo
  {journal} {Phys. Rev. X}\ }\textbf {\bibinfo {volume} {14}},\ \bibinfo
  {pages} {011012} (\bibinfo {year} {2024})}\BibitemShut {NoStop}%
\bibitem [{\citenamefont {Soriani}\ \emph {et~al.}(2025)\citenamefont
  {Soriani}, \citenamefont {Tjhung}, \citenamefont {Étienne Fodor},\ and\
  \citenamefont {Markovich}}]{soriani2025}%
  \BibitemOpen
  \bibfield  {author} {\bibinfo {author} {\bibfnamefont {Artur}\ \bibnamefont
  {Soriani}}, \bibinfo {author} {\bibfnamefont {Elsen}\ \bibnamefont {Tjhung}},
  \bibinfo {author} {\bibnamefont {Étienne Fodor}}, \ and\ \bibinfo {author}
  {\bibfnamefont {Tomer}\ \bibnamefont {Markovich}},\ }\href
  {https://arxiv.org/abs/2504.19285} {\enquote {\bibinfo {title} {Control of
  active field theories at minimal dissipation},}\ } (\bibinfo {year} {2025}),\
  \Eprint {http://arxiv.org/abs/2504.19285} {arXiv:2504.19285
  [cond-mat.stat-mech]} \BibitemShut {NoStop}%
\bibitem [{\citenamefont {Krishnan}\ \emph {et~al.}(2026)\citenamefont
  {Krishnan}, \citenamefont {Sinha},\ and\ \citenamefont
  {Mahadevan}}]{krishnan2024}%
  \BibitemOpen
  \bibfield  {author} {\bibinfo {author} {\bibfnamefont {Vishaal}\ \bibnamefont
  {Krishnan}}, \bibinfo {author} {\bibfnamefont {Sumit}\ \bibnamefont {Sinha}},
  \ and\ \bibinfo {author} {\bibfnamefont {L.}~\bibnamefont {Mahadevan}},\
  }\bibfield  {title} {\enquote {\bibinfo {title} {Hamiltonian bridge for
  scale-dependent optimal control of particles and fields},}\ }\href {\doibase
  10.1016/j.newton.2025.100365} {\bibfield  {journal} {\bibinfo  {journal}
  {Newton}\ }\textbf {\bibinfo {volume} {2}},\ \bibinfo {pages} {100365}
  (\bibinfo {year} {2026})}\BibitemShut {NoStop}%
\bibitem [{\citenamefont {Garcia-Millan}\ \emph {et~al.}(2025)\citenamefont
  {Garcia-Millan}, \citenamefont {Sch\"uttler}, \citenamefont {Cates},\ and\
  \citenamefont {Loos}}]{fbgp-qpvv}%
  \BibitemOpen
  \bibfield  {author} {\bibinfo {author} {\bibfnamefont {Rosalba}\ \bibnamefont
  {Garcia-Millan}}, \bibinfo {author} {\bibfnamefont {Janik}\ \bibnamefont
  {Sch\"uttler}}, \bibinfo {author} {\bibfnamefont {Michael~E.}\ \bibnamefont
  {Cates}}, \ and\ \bibinfo {author} {\bibfnamefont {Sarah A.~M.}\ \bibnamefont
  {Loos}},\ }\bibfield  {title} {\enquote {\bibinfo {title} {Optimal
  closed-loop control of active particles and a minimal information engine},}\
  }\href {\doibase 10.1103/fbgp-qpvv} {\bibfield  {journal} {\bibinfo
  {journal} {Phys. Rev. Lett.}\ }\textbf {\bibinfo {volume} {135}},\ \bibinfo
  {pages} {088301} (\bibinfo {year} {2025})}\BibitemShut {NoStop}%
\bibitem [{\citenamefont {Alvarado}\ \emph {et~al.}(2026)\citenamefont
  {Alvarado}, \citenamefont {Teich}, \citenamefont {Sivak},\ and\ \citenamefont
  {Bechhoefer}}]{alvarado2025}%
  \BibitemOpen
  \bibfield  {author} {\bibinfo {author} {\bibfnamefont {José}\ \bibnamefont
  {Alvarado}}, \bibinfo {author} {\bibfnamefont {Erin~G.}\ \bibnamefont
  {Teich}}, \bibinfo {author} {\bibfnamefont {David~A.}\ \bibnamefont {Sivak}},
  \ and\ \bibinfo {author} {\bibfnamefont {John}\ \bibnamefont {Bechhoefer}},\
  }\bibfield  {title} {\enquote {\bibinfo {title} {Optimal control in soft and
  active matter},}\ }\href {\doibase
  https://doi.org/10.1146/annurev-conmatphys-031324-031350} {\bibfield
  {journal} {\bibinfo  {journal} {Annu. Rev. Condens. Matter Phys.}\ }\textbf
  {\bibinfo {volume} {17}},\ \bibinfo {pages} {327--348} (\bibinfo {year}
  {2026})}\BibitemShut {NoStop}%
\bibitem [{\citenamefont {Ghosh}\ \emph {et~al.}(2024)\citenamefont {Ghosh},
  \citenamefont {Joshi}, \citenamefont {Baskaran},\ and\ \citenamefont
  {Hagan}}]{D4SM00547C}%
  \BibitemOpen
  \bibfield  {author} {\bibinfo {author} {\bibfnamefont {Saptorshi}\
  \bibnamefont {Ghosh}}, \bibinfo {author} {\bibfnamefont {Chaitanya}\
  \bibnamefont {Joshi}}, \bibinfo {author} {\bibfnamefont {Aparna}\
  \bibnamefont {Baskaran}}, \ and\ \bibinfo {author} {\bibfnamefont
  {Michael~F.}\ \bibnamefont {Hagan}},\ }\bibfield  {title} {\enquote {\bibinfo
  {title} {Spatiotemporal control of structure and dynamics in a polar active
  fluid},}\ }\href {\doibase 10.1039/D4SM00547C} {\bibfield  {journal}
  {\bibinfo  {journal} {Soft Matter}\ }\textbf {\bibinfo {volume} {20}},\
  \bibinfo {pages} {7059--7071} (\bibinfo {year} {2024})}\BibitemShut {NoStop}%
\bibitem [{\citenamefont {Alert}\ \emph {et~al.}(2022)\citenamefont {Alert},
  \citenamefont {Casademunt},\ and\ \citenamefont
  {Joanny}}]{active-turbulence-alert}%
  \BibitemOpen
  \bibfield  {author} {\bibinfo {author} {\bibfnamefont {Ricard}\ \bibnamefont
  {Alert}}, \bibinfo {author} {\bibfnamefont {Jaume}\ \bibnamefont
  {Casademunt}}, \ and\ \bibinfo {author} {\bibfnamefont {Jean-Fran\c{c}ois}\
  \bibnamefont {Joanny}},\ }\bibfield  {title} {\enquote {\bibinfo {title}
  {Active turbulence},}\ }\href {\doibase
  10.1146/annurev-conmatphys-082321-035957} {\bibfield  {journal} {\bibinfo
  {journal} {Annu. Rev. Condens. Matter Phys.}\ }\textbf {\bibinfo {volume}
  {13}},\ \bibinfo {pages} {143--170} (\bibinfo {year} {2022})}\BibitemShut
  {NoStop}%
\bibitem [{\citenamefont {Thampi}\ and\ \citenamefont
  {Yeomans}(2016)}]{active-turbulence-yeomans}%
  \BibitemOpen
  \bibfield  {author} {\bibinfo {author} {\bibfnamefont {S.~P.}\ \bibnamefont
  {Thampi}}\ and\ \bibinfo {author} {\bibfnamefont {J.~M.}\ \bibnamefont
  {Yeomans}},\ }\bibfield  {title} {\enquote {\bibinfo {title} {Active
  turbulence in active nematics},}\ }\href {\doibase
  10.1140/epjst/e2015-50324-3} {\bibfield  {journal} {\bibinfo  {journal} {Eur.
  Phys. J. Special Topics}\ }\textbf {\bibinfo {volume} {225}},\ \bibinfo
  {pages} {651--662} (\bibinfo {year} {2016})}\BibitemShut {NoStop}%
\bibitem [{\citenamefont {Shankar}\ \emph {et~al.}(2024)\citenamefont
  {Shankar}, \citenamefont {Scharrer}, \citenamefont {Bowick},\ and\
  \citenamefont {Marchetti}}]{Shankar2024}%
  \BibitemOpen
  \bibfield  {author} {\bibinfo {author} {\bibfnamefont {Suraj}\ \bibnamefont
  {Shankar}}, \bibinfo {author} {\bibfnamefont {Luca V.~D.}\ \bibnamefont
  {Scharrer}}, \bibinfo {author} {\bibfnamefont {Mark~J.}\ \bibnamefont
  {Bowick}}, \ and\ \bibinfo {author} {\bibfnamefont {M.~Cristina}\
  \bibnamefont {Marchetti}},\ }\bibfield  {title} {\enquote {\bibinfo {title}
  {Design rules for controlling active topological defects},}\ }\href {\doibase
  10.1073/pnas.2400933121} {\bibfield  {journal} {\bibinfo  {journal} {Proc.
  Natl. Acad. Sci. USA}\ }\textbf {\bibinfo {volume} {121}},\ \bibinfo {pages}
  {e2400933121} (\bibinfo {year} {2024})}\BibitemShut {NoStop}%
\bibitem [{\citenamefont {Radhakrishnan}\ \emph {et~al.}(2026)\citenamefont
  {Radhakrishnan}, \citenamefont {Serafin}, \citenamefont {Schmidt},\ and\
  \citenamefont {Fodor}}]{radhakrishnan2025}%
  \BibitemOpen
  \bibfield  {author} {\bibinfo {author} {\bibfnamefont {Byjesh~N}\
  \bibnamefont {Radhakrishnan}}, \bibinfo {author} {\bibfnamefont {Francesco}\
  \bibnamefont {Serafin}}, \bibinfo {author} {\bibfnamefont {Thomas~L}\
  \bibnamefont {Schmidt}}, \ and\ \bibinfo {author} {\bibfnamefont {Étienne}\
  \bibnamefont {Fodor}},\ }\bibfield  {title} {\enquote {\bibinfo {title}
  {Irreversibility in scalar active turbulence: the role of topological
  defects},}\ }\href {\doibase 10.1088/1367-2630/ae44c3} {\bibfield  {journal}
  {\bibinfo  {journal} {New J. Phys.}\ }\textbf {\bibinfo {volume} {28}},\
  \bibinfo {pages} {034601} (\bibinfo {year} {2026})}\BibitemShut {NoStop}%
\end{thebibliography}%
\end{document}